\documentclass[a4paper,11pt]{article}
\pdfoutput=1 

\usepackage{jheppub} 
\usepackage[bottom]{footmisc}
\usepackage{amssymb}
\usepackage{amsmath}
\usepackage{amsthm}
\usepackage[usenames,dvipsnames]{xcolor}
\usepackage{epsfig}
\usepackage{dcolumn}
\usepackage{tikz}
\usetikzlibrary{shapes.geometric, arrows,positioning}
\usepackage{upgreek}
\usepackage{setspace}
\usepackage{array,multirow,bigdelim,arydshln}
\usepackage{appendix}
\usepackage{xparse}
\usepackage{stmaryrd}
\usepackage[T1]{fontenc} 
\usepackage{mathtools}
\usepackage{physics} 
\usepackage{adjustbox}
\usepackage{multirow}
\usepackage{graphicx} 
\usepackage{subcaption}
\usepackage{float} 
\graphicspath{{./images/}}
\usepackage{comment}
\usepackage[nottoc]{tocbibind}
\usepackage{hyperref}
\usepackage[utf8]{inputenc}
\usepackage{CJK}
\hypersetup{
	colorlinks,
	urlcolor=Maroon,
	linkcolor=Maroon,
	citecolor=Maroon
	}

\NewDocumentCommand{\binomial}{omm}
 {%
  \genfrac(){0pt}{}{#2}{#3}%
  \IfValueT{#1}{_{\!#1}}%
 }
\NewDocumentCommand{\eulerian}{omm}
 {%
  \genfrac<>{0pt}{}{#2}{#3}%
  \IfValueT{#1}{_{\!#1}}%
 }

\usepackage{latexsym}
\usepackage{tikz}

\theoremstyle{plain}

\newtheorem{thm}{Theorem}[section]

\theoremstyle{definition}
\newtheorem{example}[thm]{Example}

\def\bea#1\eea{\begin{eqnarray}#1\end{eqnarray}}
\def\be#1\ee{\begin{equation}#1\end{equation}}
\def\ba#1\ea{\begin{align}#1\end{align}}

\usepackage{amsmath}
\usepackage{multicol}
\usepackage{bbm}
\usepackage{enumerate}

\usepackage{amsthm}
\usepackage{mathrsfs}
\usepackage{upgreek}
\usepackage{amssymb}
\usepackage{bm}
\usepackage{setspace}
\usepackage{array,multirow,arydshln}
\usepackage{bigdelim}
\usepackage{scalerel}
\usepackage{diagbox}

\usepackage{tabularx}

\usepackage{tikz}

\usetikzlibrary{shapes.geometric,arrows,arrows.meta,decorations.pathmorphing,decorations.markings,patterns}

\def\<{\langle}
\def\>{\rangle}

\def\Tr{\text {Tr}}

\usepackage[percent]{overpic}
\usepackage{multirow} 
\usepackage{slashed}

\title{On differential operators for scalar-scaffolded gluons}

\author[b]{Jin Dong,}
\author[c]{Yong-Xiang Su}
\author[a,d,e]{and Dongyu Yang}

\affiliation[a]{School of Fundamental Physics and Mathematical Sciences, Hangzhou Institute for Advanced Study, UCAS, Hangzhou 310024, China}
\affiliation[b]{Max-Planck-Institut f\"ur Physik, Werner-Heisenberg-Institut, Boltzmannstr. 8, 85748 Garching bei M\"unchen, Germany}

\affiliation[c]{School of Physical Sciences, University of Science and Technology of China, Hefei, Anhui 230026, China}

\affiliation[d]{Institute of Theoretical Physics, Chinese Academy of Sciences, Beijing 100190, China}
\affiliation[e]{School of Physical Sciences, University of Chinese Academy of Sciences, Beijing 100049, China}

\emailAdd{jindong@mpp.mpg.de}
\emailAdd{anonym@mail.ustc.edu.cn}
\emailAdd{yangdongyu24@mails.ucas.ac.cn}

\abstract{
Recently, based on the curve-integral formulation for stringy Tr$\phi^3$ amplitudes, a combinatorial formulation for Yang-Mills amplitudes has been proposed which describes gluons using pairs of scalars and produces the $n$-gluon amplitude from simple kinematical shift of stringy Tr$\phi^3$ amplitudes with $2n$ scalars. It has revealed a variety of new properties and structures even for tree-level gluon amplitudes such as hidden zeros and splits, and in this note we provide another example: we study differential operators acting on Yang-Mills amplitudes with respect to $2n$-scalar kinematic variables, which convert such scalar-scaffolded gluons into scalars. In particular, we find $(n{-}1)$-fold differential operators (using $2n$-scalar variables) that turn the $n$-gluon amplitude into a single planar $\phi^3$ diagram; we then generalize such operators to those that convert $n$ gluons to mixed amplitudes with $r$ scalars and $n{-}r$ gluons (the latter can be viewed as insertions on $\phi^3$ diagrams). We also show that the number of linearly independent mixed amplitudes with $r$ scalars and $n-r$ gluons is given by the number of $\phi^3$ diagrams, the Catalan number $\mathcal{C}_{r-2}$, which can be viewed as a generalization of the ``uniqueness'' theorem of gluon amplitudes (with $r=0$). Finally, our construction leads to a planar version of the universal expansion of Yang-Mills amplitudes into a sum of gauge-invariant prefactors built from nested commutators, each accompanied by a mixed amplitude in the natural basis. This formulation significantly reduces the redundancy present in the original expansion.

}
\preprint{MPP-2025-229}

\begin{document}

\begin{CJK*}{UTF8}{}
\CJKfamily{gbsn}
\maketitle
\end{CJK*}
\addtocontents{toc}{\protect\setcounter{tocdepth}{2}}

\numberwithin{equation}{section}

		\tikzset{
		particles/.style={dashed, postaction={decorate},
			decoration={markings,mark=at position .5 with {\arrow[scale=1.5]{>}}
		}}
	}
	\tikzset{
		particle/.style={draw=black, postaction={decorate},
			decoration={markings,mark=at position .5 with {\arrow[scale=1.1]{>}}
		}}
	}
	\def  \layersep {.6cm}

\section{Introduction}
Over the past few decades, significant progress has been made in uncovering the hidden structures of scattering amplitudes in quantum field theory. Perhaps one of the most celebrated developments is the Bern-Carrasco-Johansson (BCJ) double copy relation~\cite{Bern:2008qj, Bern:2010ue}, which establishes a deep connection between gauge and gravity amplitudes. The Cachazo-He-Yuan (CHY) formula~\cite{Cachazo:2013gna,Cachazo:2013hca,Cachazo:2013iea} has further unified a broad class of theories into a common foundation, enabling explicit constructions of Yang-Mills (YM) amplitudes and their gravitational double copies~\cite{Mafra:2011kj,Fu:2017uzt, Du:2017kpo, Teng:2017tbo} (see also~\cite{Cheung:2020tqz, Edison:2020ehu, He:2021lro, Wei:2023yfy, Chen:2024gkj, Gomez:2025tqx} for recent advances). Concretely, in the Del Duca-Dixon-Maltoni (DDM) half-ladder basis~\cite{DelDuca:1999rs}, the YM amplitude can be expressed as a sum over $(n-2)!$ bi-adjoint $\phi^3$ amplitudes accompanied by the corresponding BCJ numerators. The CHY representation makes this structure manifest and provides a powerful tool for computing explicit  BCJ numerators. However, their naive construction  depends on a reference ordering, which breaks crossing symmetry and complicates the computation (one solution is to restore the crossing symmetry as in~\cite{Edison:2020ehu}, which is equivalent to averaging over all reference orderings). On the other hand, if one is only concerned with color-ordered YM amplitudes, the asymptotic complexity is expected to grow exponentially, since the number of ordered cubic diagrams is given by the Catalan number.

A recent breakthrough on reformulation of scattering amplitudes for $\operatorname{Tr}\phi^3$ to all orders in the `t Hooft topological expansion, known as the curve integral has been introduced in~\cite{Arkani-Hamed:2023lbd,Arkani-Hamed:2023mvg}. This framework has also been generalized to the scattering of pions and gluons in the planar limit in a series of works~\cite{Arkani-Hamed:2023swr,Arkani-Hamed:2023jry,Arkani-Hamed:2024nhp} revealing hidden zeros of corresponding amplitudes as well as novel properties such as factorization near zeros and special splitting behaviors~\cite{Arkani-Hamed:2023swr, Cao:2024gln, Bartsch:2024amu, Li:2024qfp,Arkani-Hamed:2024fyd} (see also for related results~\cite{Rodina:2024yfc, Cao:2024qpp, Zhou:2024ddy, Li:2024bwq, GimenezUmbert:2024jjn, Zhang:2024efe, Zhang:2024iun, De:2025bmf, Backus:2025hpn, Chang:2025cqe, Feng:2025ofq, Feng:2025dci, Zhang:2025zjx,Zhou:2025tvq}). With this approach, tree-level YM amplitudes (and even loop integrands) admit a novel representation in terms of scalar-scaffolding variables. These variables trivialize the constraint of momentum conservation and provide a simple, unique representation of YM amplitudes, while key properties such as gauge invariance and factorization at simple poles are presented in a different way. The above development suggests that there might be a better approach to the YM amplitudes, of which the complexity grows exponentially. 

On the other hand, in the traditional representation, differential operators involving polarization vectors and momenta relate YM amplitudes to both YM+$\phi^3$ mixed amplitudes~\cite{Cheung:2017ems} and pure scalar ones by systematically lowering the spin of external states. The products of these operators act locally on the kinematic data and preserve essential structural properties, {\it i.e.} momentum conservation and gauge invariance, making them an efficient tool for navigating between theories of different spin. Nevertheless, the analogous story for scalar-scaffolded amplitudes remains largely unexplored (however, see~\cite{Backus:2025njt} for a related paper). Since the gluon, scalar, and mixed amplitudes are all simple and unique objects in these variables, it is natural to expect a straightforward connection through differential operators. Establishing such relations would not only generalize the traditional transmutation program into the new formalism, but also shed light on hidden structures of YM amplitudes and potentially offer a more efficient route for generating mixed amplitudes within the planar framework.

In this paper, we begin to address the above questions through the following steps. We first develop a systematic method for deriving differential operators that extract single scalar cubic diagrams from the scalar-scaffolded YM amplitude. Building on this construction, we then investigate generalizations of these differential operators, showing that while in some cases they naturally lead to insertions of gluons, in other cases the naive generalization fails. We also demonstrate that the number of linearly independent mixed amplitudes of $r$ scalars and $n-r$ gluons is counted by the Catalan number $\mathcal{C}_{r-2}$. Altogether, these developments lead to a planar version of the universal expansion of Yang-Mills amplitudes, expressed as a sum of gauge-invariant prefactors built from nested commutators, each paired with a mixed amplitude in the natural basis. This result offers a streamlined perspective that greatly reduces the redundancy of the traditional expansion.

This paper is organized as follow: In section \ref{sec: review}, we review the scalar-scaffolded gluon amplitudes and transmutation operators, as well as the universal expansion of YM amplitudes into YM$+\phi^3$ mixed amplitudes and pure scalar ones. In section \ref{sec: derivative to cubic}, we present a systematic algorithm for obtaining single $\phi^3$ diagram by operating a sequence of derivatives acting on the scalar-scaffolded YM amplitude and provide proofs of our rules for certain cases using factorization. Based on the insight from planarity, we show in section \ref{sec: minimal basis} that the number of independent mixed amplitudes is determined only by the number of scalars within, counted by Catalan numbers. Then in section  \ref{sec: gluon insertion}, we discuss the feasibility that whether similar differential operators could be used to derived mixed amplitudes within the natural planar basis. Finally section \ref{sec: universal expansion} reorganizes the conventional form of universal expansion into such basis.

\section{Review}\label{sec: review}
\subsection{Scaffolded gluon amplitudes}
\begin{figure}
    \centering
    \includegraphics[scale=1.2]{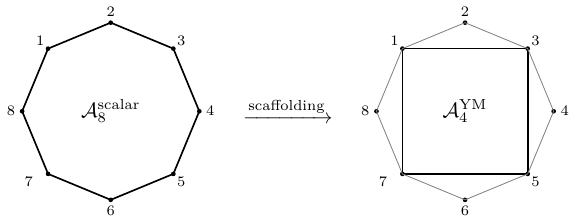}
    \caption{Scaffolding from 8-scalar amplitude $\mathcal{A}_8^{\text{scalar}}(1,2,3,4,5,6,7,8)$ to 4-gluon amplitude $\mathcal{A}_4^{\rm YM}(1,3,5,7)$.}
    \label{fig:scaffolding}
\end{figure}

In this section, we give a lightning review of scalar-scaffolded gluon amplitudes\cite{Arkani-Hamed:2023swr,Arkani-Hamed:2023jry,Arkani-Hamed:2024tzl}~\footnote{Similar ideas were also briefly discussed in previous works; see for example ~\cite{Arkani-Hamed:2008owk,Stieberger:2014cea,Stieberger:2015qja, Cachazo:2015ksa}.}. To begin with, we define the planar variables
\begin{equation}
    X_{i,j}\equiv(p_i+p_{i+1}+\cdots+p_{j-1})^2.
\end{equation}
By definition, we have $X_{i,j}=X_{j,i}$ and $X_{i,i}=X_{i,i+1}=0$ for massless particles. The usual Mandelstam variables are related to these invariants via the ABHY~\cite{Arkani-Hamed:2017mur} conditions
\begin{equation}
    c_{i,j}\equiv-2p_i\cdot p_j=X_{i,j}+X_{i+1,j+1}-X_{i,j+1}-X_{i+1,j}.
\end{equation}
For an $n$-point gluon amplitude, we consider expanding the external gluon legs into double scalar legs with the following Feynman vertex:
\begin{equation}
\vcenter{\hbox{\includegraphics[scale=1.2]{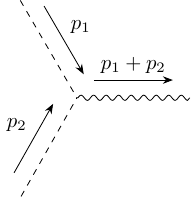}}}=g_{\mathrm{YM}}(p_2-p_1)^\mu,
\end{equation}
where $g_{\mathrm{YM}}$ is the gluon coupling constant. 
In this setup, each pair of YMS scalars is scaffolded into a gluon. To extract the $n$-point YM amplitude, one must take the intermediate gluons on-shell in the $2n$-point ({\it i.e.} $n$-pair) amplitude, evaluated on the following scaffolding residues:
\begin{equation}
    X_{1,3}=X_{3,5}=X_{5,7}=\cdots=X_{2n-1,1}=0.
\end{equation}
Such operation can be depicted on a surface with $2n$-punctures on the edge. Taking the scaffolding of a scalar amplitude is equivalent pictorially as pinching the $2i-1$-th and $2i$-th vertex into one and collapse the $2n$-gon into an $n$-gon (see figure \ref{fig:scaffolding}).

It is straightforward to verify that the kinematic data, specifically the gluon polarizations and momenta, are fully captured by the scaffolded planar variables. For example, for an tree-level $n$-point Yang-Mills amplitude, there are $n(n-3)/2$ independent scalar products $k_i \cdot k_j$, $n(n-1)/2$ polarization contractions $\epsilon_i \cdot \epsilon_j$, and $n(n-1) - n$ mixed terms $\epsilon_i \cdot p_j$, giving a total of $2n(n-2)$ independent kinematic variables. On the other hand, for a $2n$-point scalar amplitude, the number of independent kinematic variables is $2n(2n-3)/2 - n = 2n(n-2)$, which precisely matches the count for the gluon case.

To be precise, the momentum of the $i$-th gluon $k_i^\mu$ is given by 
\begin{equation}\label{eq:momenta to 2n-gon}
    k_i^\mu=(p_{2i-1}+p_{2i})^\mu.
\end{equation}
The polarization, on the other hand, contains a free parameter $\lambda$, which is associated with the gauge redundancy.
\begin{equation}\label{eq:polarization to 2n-gon}
    \epsilon_i^\mu=(1-\lambda)p_{2i-1}^\mu-\lambda p_{2i}^\mu,
\end{equation}
which one can verify it satisfying the transverality condition $\epsilon_i\cdot k_i=0$ and manifesting a generalized ``circular polarizations'' $\epsilon_i\cdot\epsilon_i=0$.

A benefit for the scaffolding language is that it can manifest the gauge invariance and multi-linearity in a unifying way. For the $i$-th gluon, its gauge transformation reads
\begin{equation}
    \epsilon_i^\mu\to\epsilon_i^\mu+\alpha k_i^\mu\qquad\forall\alpha\in\mathbb C,
\end{equation}
which can be translated into (with $\lambda=\frac{1}{2}$)
\begin{equation}
    X_{2i,j}\to X_{2i,j}+\alpha(X_{2i+1,j}-X_{2i-1,j}).
\end{equation}
Thus the gauge invariance condition for the $i$-th gluon becomes
\begin{equation}\label{eq:gauge invariance}
    \mathcal{A}_n^{\rm YM}[ X_{2i,j}+\alpha(X_{2i+1,j}-X_{2i-1,j})]-\mathcal{A}_n^{\rm YM}[X_{2i,j}]=0.
\end{equation}
For the multi-linearity, consider
\begin{equation}
    \epsilon_i^\mu\to\epsilon_i^\mu+\beta\epsilon_i^\mu\qquad\forall\beta\in\mathbb C,
\end{equation}
which is given by
\begin{equation}
    X_{2i,j}\to X_{2i,j}+\frac{\beta}{2}(X_{2i,j}-X_{2i-1,j})+\frac{\beta}{2}(X_{2i,j}-X_{2i+1,j}).
\end{equation}
This gives us the scaffolding form of multi-linearity condition on $i$-th gluon
\begin{equation}\label{eq:multi-linearity}
    \mathcal{A}_n^{\rm YM}\left[X_{2i,j}\to X_{2i,j}+\frac{\beta}{2}(X_{2i,j}-X_{2i-1,j})+\frac{\beta}{2}(X_{2i,j}-X_{2i+1,j})\right]-\mathcal{A}_n^{\rm YM}[X_{2i,j}]=\beta\mathcal{A}_n^{\rm YM}[X_{2i,j}].
\end{equation}
Combine equations \eqref{eq:gauge invariance} and \eqref{eq:multi-linearity}, we obtain a compact formula
\begin{multline}
    \mathcal{A}_n^{\rm YM}\left[X_{2i,j}\to X_{2i,j}+\alpha(X_{2i+1,j}-X_{2i-1,j})+\frac{\beta}{2}(X_{2i,j}-X_{2i-1,j})+\frac{\beta}{2}(X_{2i,j}-X_{2i+1,j})\right]\\-\mathcal{A}_n^{\rm YM}[X_{2i,j}]=\beta\mathcal{A}_n^{\rm YM}[X_{2i,j}].
\end{multline}
Picking $\alpha=\pm\frac{\beta}{2}$, we have:
\begin{equation}
    \mathcal{A}_n^{\rm YM}[X_{2i,j}\to X_{2i,j}+\beta(X_{2i,j}-X_{2i-1,j})] -\mathcal{A}_n^{\rm YM} =\beta \mathcal{A}_n^{\rm YM},
\end{equation}
\begin{equation}
    \mathcal{A}_n^{\rm YM}[X_{2i,j}\to X_{2i,j}+\beta(X_{2i,j}-X_{2i+1,j})] -\mathcal{A}_n^{\rm YM} =\beta \mathcal{A}_n^{\rm YM}.
\end{equation}
This is equivalent to the following:
\begin{equation} \label{eq:gauge1}
    \mathcal{A}_n^{\rm YM}= \sum_{j\neq i-1,i,i+1} (X_{2i,j}-X_{2i-1,j}) \frac{\partial}{\partial X_{2i,j}} \mathcal{A}_n^{\rm YM}\,,
\end{equation}
\begin{equation} \label{eq:gauge2}
    \mathcal{A}_n^{\rm YM}= \sum_{j\neq i-1,i,i+1} (X_{2i,j}-X_{2i+1,j}) \frac{\partial}{\partial X_{2i,j}} \mathcal{A}_n^{\rm YM}\,,
\end{equation}
where we have used the fact that at each given term, even label $2i$ appears as a subscript in at most one $X$. Note that one of the above conditions can be replaced by: 
\begin{equation} \label{eq:gauge3}
   \eqref{eq:gauge1}-\eqref{eq:gauge2}= \sum_{j\neq i-1,i,i+1} (X_{2i+1,j}-X_{2i-1,j}) \frac{\partial}{\partial X_{2i,j}} \mathcal{A}_n^{\rm YM}=0\,.
\end{equation}

\begin{figure}
    \centering
    \includegraphics[scale=1.1]{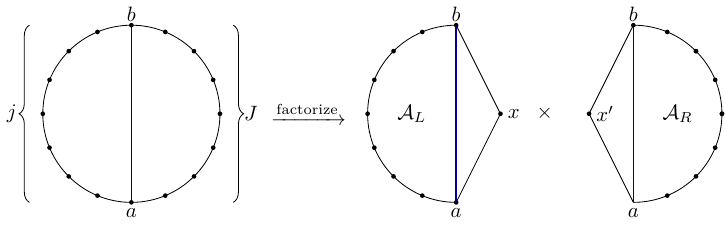}
    \caption{ $\mathcal{A}_n^{\rm YM}$ factorizes on $X_{a,b}$, forming a pair of sub-amplitudes $\mathcal{A}_L$ and $\mathcal{A}_R$.}
    \label{fig:factorization}
\end{figure}

The factorization of the gluon amplitude also manifests in our scalar-scaffolded language. Denote $X_{a,b}$ for both $a,b$ odd as the gluon propagator we are considering, the residue of the gluon amplitude at $X_{a,b}=0$ is given by the linear combination of planar variables and 
derivative of partial amplitudes:
\begin{equation} \label{eq:fac1}
    {\rm Res}_{X_{a,b}=0} \mathcal{A}_n^{\rm YM}= \sum_{j,J} (X_{j,J}-X_{a,j}-X_{b,J}) \frac{\partial }{\partial X_{x,j}} \mathcal{A}_L \, \frac{\partial }{\partial X_{x',J}} \mathcal{A}_R,
\end{equation}
where the summation is over all $j=a+1,\dots,b-1$ and $J=b+1,\dots,a-1$, and punctures $x$ and $x'$ are axillary points to ensure that the partial amplitudes $\mathcal{A}_L(a,\dots,b,x)$ and $\mathcal{A}_R(b,\dots,x')$ are also formulated by scaffolded scalars (see figure \ref{fig:factorization}). 

Given that one can label the vertices either clockwise or counterclockwise, it is natural to expect a greater symmetry in the labels $a$ and $b$ appearing in the prefactor. In fact, there are three additional equivalent ways to express the factorization, differing only in the subscripts $a$ and $b$ of the prefactor. Let us list them explicitly:

\begin{equation} \label{eq:fac2}
    {\rm Res}_{X_{a,b}=0} \mathcal{A}_n^{\rm YM}= \sum_{j,J} (X_{j,J}-X_{b,j}-X_{a,J}) \frac{\partial }{\partial X_{x,j}} \mathcal{A}_L \, \frac{\partial }{\partial X_{x',J}} \mathcal{A}_R\,,
\end{equation}

\begin{equation} \label{eq:fac3}
    {\rm Res}_{X_{a,b}=0} \mathcal{A}_n^{\rm YM}= \sum_{j,J} (X_{j,J}-X_{a,j}-X_{a,J}) \frac{\partial }{\partial X_{x,j}} \mathcal{A}_L \, \frac{\partial }{\partial X_{x',J}} \mathcal{A}_R\,,
\end{equation}

\begin{equation} \label{eq:fac4}
    {\rm Res}_{X_{a,b}=0} \mathcal{A}_n^{\rm YM}= \sum_{j,J} (X_{j,J}-X_{b,j}-X_{b,J}) \frac{\partial }{\partial X_{x,j}} \mathcal{A}_L \, \frac{\partial }{\partial X_{x',J}} \mathcal{A}_R\,.
\end{equation}
It is easy to verify that the difference between the right hand sides of any two of these expressions vanishes. For example, the difference between \eqref{eq:fac1} and \eqref{eq:fac3} is given by
\begin{equation} 
     \sum_j \frac{\partial }{\partial X_{x,j}} \mathcal{A}_L \sum_{J} (X_{a,J}-X_{b,J})  \, \frac{\partial }{\partial X_{x',J}} \mathcal{A}_R\,, 
\end{equation}
where we have
\begin{equation} \label{eq:fac_dif}
     (X_{a,J}-X_{b,J})  \, \frac{\partial }{\partial X_{x',J}} \mathcal{A}_R=0 \,.
\end{equation}
Here we have used the fact that in $\mathcal{A}_R$ we have labeled $x'-1 = a$ and $x'+1 = b$, therefore the expression \eqref{eq:fac_dif} matches exactly the identity~\eqref{eq:gauge3} when applied to $\mathcal{A}_R$.

To conclude this section, we present the translation between the usual Lorentz products and the scalar-scaffolded planar variables~\cite{Arkani-Hamed:2023jry}. For simplicity, we set $\lambda = 1$, which does not affect the gauge-invariant amplitude:
\begin{align} \label{eq: epXtrans}
&2 \epsilon_i \cdot \epsilon_j=X_{2 i,2 j-1}+ X_{2 i-1,2 j}- X_{2 i-1,2 j-1}-X_{2 i,2 j}\,, \nonumber  \\ 
&2 \epsilon_i \cdot k_j=X_{2i, 2j{+}1}-X_{2i, 2j{-}1}+X_{2i{-}1, 2j{-}1}-X_{2i{-}1, 2j{+}1} \,, \\ 
 &2 k_i \cdot k_j= X_{2 i-1,2 j+1}+X_{2 i+1,2 j-1}- X_{2 i-1,2 j-1} - X_{2 i+1,2 j+1}. \nonumber
\end{align}
For example, the explicit expression for $n=3$ reads:
\begin{equation}
    \mathcal{A}_3^{\rm YM}=-4 (k_2\cdot \epsilon_3\epsilon_1\cdot\epsilon_2+k_1\cdot \epsilon_2\epsilon_1\cdot\epsilon_3+k_2\cdot\epsilon_1\epsilon_2\cdot\epsilon_3).
\end{equation}
Substituting equations~\eqref{eq: epXtrans} the amplitude becomes
\begin{equation} \label{eq:YM3pt}
    \mathcal{A}_3^{\rm YM}=-X_{1,4}X_{2,5}+X_{1,4}X_{2,6}-X_{1,4}X_{3,6}+X_{2,4}X_{3,6}-X_{2,5}X_{3,6}+X_{2,5}X_{4,6}.
\end{equation}
For explicit results up to $n=7$, see~\cite{Arkani-Hamed:2024tzl}. We also employ the $2n$ scalar-scaffolded variables to describe the mixed amplitudes of gluons and colored $\phi^3$ scalars~\cite{Cachazo:2014xea}. Note that an even label $2i$ appears in the subscripts of the $X$'s only if particle $i$ is a gluon. For example,
\begin{equation}
    \mathcal{A}_3^{\mathrm{YM}+\phi^3}(\{2\}|1,3)= -2 k_1\cdot \epsilon_2=X_{1,4}\,,
\end{equation}
where we consider the three-point mixed amplitude with particle $2$ being a gluon; hence the even label $4$ appears.

\subsection{Transmuted operators} \label{sec: trans operators}

An elegant way to relate scattering amplitudes across different theories is through the use of specific differential operators~\cite{Cheung:2017ems,Cheung:2017yef}. For our purposes, we focus on those operators that transform gluons into bi-adjoint scalars. To be more explicitly, these operators are defined by a set of derivatives on kinematic variables:
\begin{align}
    \mathcal{T}_{i,j}&:=\partial_{\epsilon_i\cdot\epsilon_j},\\
    \mathcal{T}_{i,j,k}&:=\partial_{p_i\cdot\epsilon_j}-\partial_{p_k\cdot\epsilon_j}.
\end{align}
Crucially, any well-defined combination of these operators must preserve both momentum conservation and gauge invariance. To incorporate these constraints we define a the total momentum operator for momentum conservation
\begin{equation}
    \mathcal{P}_v:=\sum_{i=1}^np_i\cdot v,
\end{equation}
and the Ward identity operator for gauge invariance
\begin{equation}
    \mathcal{W}_i:=\sum_vp_i\cdot v\partial_{v\cdot \epsilon_i},
\end{equation}
where $i$ denotes external legs and $v$ labels any momentum or polarization vector. 
Any gauge invariant and momentum conserving amplitude $A$ is annihilated by these operators, that is to say
\begin{equation}
   \mathcal P_v\cdot A=\mathcal{W}_i\cdot A=0.
\end{equation}
One can check that both length-2 and length-3 transmuted operators satisfy momentum conservation
\begin{equation}
    [\mathcal{P}_v,\mathcal{T}_{i,j}]\cdot A=[\mathcal{P}_v,\mathcal{T}_{i,j,k}]\cdot A=0.
\end{equation}
However, only the length-2 transmuted operator seems to obey gauge invariance such that 
\begin{equation}
    [\mathcal{W}_l,\mathcal{T}_{i,j}]\cdot A=0,
\end{equation}
while as an exception, the commutator of $\mathcal{T}_{i,j,k}$ with $\mathcal{W}_l$ reveals a non-trivial result
\begin{equation}
    [\mathcal{W}_l,\mathcal{T}_{i,j,k}]\cdot A=\delta_{kl}\mathcal{T}_{j,k}\cdot A-\delta_{il}\mathcal{T}_{i,j}\cdot A.
\end{equation}
We emphasize that even as $\mathcal{T}_{i,j,k}$ does not manifest gauge invariance intrinsically, it can still be effectively gauge invariant when combined with enough supplemental transmuted operators. For example, the commutator of $\mathcal{W}_l$ with the product $\mathcal{T}_{i,k}\mathcal{T}_{i,j,k}$ gives
\begin{equation}
    [\mathcal{T}_{i,k}\cdot\mathcal{T}_{i,j,k},\mathcal{W}_l]\cdot A=\delta_{il}\mathcal{T}_{i,k}\cdot\mathcal{T}_{i,j}\cdot A-\delta_{k,l}\mathcal{T}_{i,k}\cdot\mathcal{T}_{j,k}\cdot A=0,
\end{equation}
where at the second equal sign we have used the condition that the amplitude $A$ is multi-linear in polarizations $\epsilon_i$ and $\epsilon_k$, leading it vanishing after being twice interacted by transmuted operators $\mathcal{T}_{i,k}\cdot\mathcal{T}_{i,j}$ and $\mathcal{T}_{i,k}\cdot\mathcal{T}_{j,k}$.

Applied in various combinations, the transmutation operators form an interlocking web of scattering
amplitudes relations across a wide range of theories. For single trace operators defined by a sequence of transmuted operators in an ordered list $\alpha$
\begin{equation}
    \mathcal{T}[\alpha]:=\mathcal{T}_{\alpha_1,\alpha_n}\cdot\prod_{i=2}^{n-1}\mathcal{T}_{\alpha_{i-1},\alpha_i,\alpha_n},
\end{equation}
they can transform an amplitude of gluons into the one of gluons mixed with bi-adjoint scalars
\begin{equation} \label{eq: ep operator}
    \mathcal{T}[\alpha]\mathcal{A}_n^{\rm YM}=\mathcal{A}_n^{\mathrm{YM}+\phi^3}(\{\bar\alpha\}|\alpha),
\end{equation}
in which $\{\bar\alpha\}$ denotes remaining gluons. 

Given the success of these operators, it is natural to ask whether analogous operators exist in the scaffolded variables acting directly on amplitudes expressed in this form. Since the relation between these variables and the usual Lorentz products was discussed in~\cite{Arkani-Hamed:2023jry}, one can in principle obtain an exact map using the chain rule. However, as we will see later, there exists a more natural choice that leads to a better basis.

\subsection{Universal expansion}
The expansion of Yang-Mills amplitudes on Yang-Mills-scalar amplitudes was originally proposed in \cite{Lam:2016tlk,Fu:2017uzt,Du:2017kpo} by studying the CHY representation of Yang-Mills and Yang-Mills scalar amplitudes, and later realized as a consequence of basic principles {\it i.e.} power counting, locality and gauge invariance \cite{Dong:2021qai}. The explicit formula is given by
\begin{equation} \label{eq:expansion}
\begin{aligned}
    {\cal A}^\text{YM}_n
    = \sum_{m,\alpha\in S_m} (-1)^m \, \epsilon_1 \cdot f_{\alpha_1} \cdot f_{\alpha_2} 
    \cdots f_{\alpha_m} \cdot \epsilon_n  \, {\cal A}_n^{\mathrm{YM}+\phi^3}(\{\bar{\alpha}\}|1,\alpha,n)\,,
\end{aligned}
\end{equation}
Note that both particle $1$ and $n$ are special and the sum is over all {\it ordered subsets} of $\{2,3,\cdots,n-1\}$ denoted by $\alpha$ (with $|\alpha|=m$ for $m=0, 1, \cdots, n{-}2$), and $\{\bar{\alpha}\}$ denotes the complementary (unordered) set with $n{-}2{-}m$ labels. For particles $2,\ldots,(n-1)$, the gauge invariance is manifest in \eqref{eq:expansion}, since we are summing over prefactors that contain $f_i^{\mu \nu} \equiv k_i^\mu \epsilon_i^\nu - \epsilon_i^\mu k_i^\nu$ multiplied with a mixed amplitude involving gluons ${\bar{\alpha}}$ and doubly ordered $\phi^3$ scalars, where the second ordering is labeled by $(1,\alpha,n)$. The gauge invariance of particle $1$(or $n$), however, is not obvious. Actually, according to \cite{Arkani-Hamed:2016rak}, if we express the Yang-Mills amplitudes with a local form, as is the case in \eqref{eq:expansion}, it is only possible to manifest the gauge invariance of $(n-2)$ particles. 
Furthermore, as proven in \cite{Dong:2021qai}, requiring the gauge invariance of $(n-2)$ particles already spits out the prefactors, yet one can not confirm that the  gauge invariant block accompanied with the prefactor is indeed a YMS amplitude. The next step is to use the gauge invariance of another particle say the first particle, which uniquely fixes the Yang-Mills amplitudes, as well as its expansion \eqref{eq:expansion}.

As to present more concretely, we provide a few examples for this expansion. For $n=3$, the expansion is quite trivial, 
\begin{equation}
    \mathcal{A}_3^{\rm YM}=\epsilon_1\cdot\epsilon_3\mathcal{A}_3^{\mathrm{YM}+\phi^3}(\{2\}|13)-\epsilon_1\cdot f_2\cdot\epsilon_3\mathcal{A}_3^{\mathrm{YM}+\phi^3}(\varnothing|123).
\end{equation}
For $n=4$, we have 3 different kinds of terms, each corresponds to $m=0,1,2$:
\begin{equation}\label{eq:4pt universal expansion}
    \begin{aligned}
        \mathcal{A}_4^{\rm YM}&=\epsilon_1\cdot\epsilon_4\mathcal{A}_4^{\mathrm{YM}+\phi^3}(\{2,3\}|1,4)-\epsilon_1\cdot f_2\cdot\epsilon_4\mathcal{A}_4^{\mathrm{YM}+\phi^3}(\{3\}|1,2,4)-\epsilon_1\cdot f_3\cdot\epsilon_4\mathcal{A}_4^{\mathrm{YM}+\phi^3}(\{2\}|1,3,4)\\
        &\ +\epsilon_1\cdot f_2\cdot f_3\cdot\epsilon_4\mathcal{A}_4^{\mathrm{YM}+\phi^3}(\varnothing|1,2,3,4)+\epsilon_1\cdot f_3\cdot f_2\cdot\epsilon_4\mathcal{A}_4^{\mathrm{YM}+\phi^3}(\varnothing|1,3,2,4),
    \end{aligned}
\end{equation}
We also provide explicit expression for 5-point expansion in Appendix~\ref{app: expansion}.

We remark that the number of terms in this expansion is
\begin{equation}
\sum_{m=0}^{n-2} m! \binom{n-2}{m} ,
\end{equation}
which equals to $2, 5, 16, 65, 326, 1957$ for $n=3,4,5,6,7,8$, and asymptotically behaves as $(n-2)!\, \mathrm{e}$ for large $n$ (see \href{https://oeis.org/A000522}{OEIS A000522}). It is then reasonable to ask whether there exists a planar analogue of this expansion in the scaffolded variables that respects the Catalan-number counting.

\section{Scalar Cubic Diagram from Derivatives}\label{sec: derivative to cubic}
In this section, we investigate differential operators that generate single cubic diagrams. Note that the derivatives involving conventional variables built from Lorentz contractions of $\epsilon$ and $p$ are highly restrictive under momentum conservation~\cite{Cheung:2017ems}. In contrast, the planar variables automatically encode momentum conservation, admit derivatives free of such constraints.

The power counting of YM amplitudes is given by $X^2$ for any $n$. To extract $\phi^3$ scalar diagrams (amplitudes) that scale as $1/X^{n-3}$, one needs to act with derivatives involving $n-1$ distinct planar variables $X_{i,j}$. Such derivatives are not expected to involve $X_{o,o}$, since these variables could appear in the denominator. Moreover, they are constrained by multi-linearity: in each term, every even label $2i$ (with $i = 1, 2, \ldots, n$) appears as a subscript in at most one $X$. Consequently, we have 
\begin{equation}
\label{eq:labelMulti}
\frac{\partial^2}{\partial X_{2a,b} \partial X_{2a,b'}} \mathcal{A}^{\rm YM}_n=0. 
\end{equation}
And we expect the relevant derivatives to involve $n-1$ variables: one of the form $X_{e,e}$ and $n-2$ of the form $X_{e,o}$, {\it i.e.} 
\begin{equation}
    \mathcal{D}_{(2a',2b'),(2a_1,2b_1-1),\ldots,(2a_{n-2},2b_{n-2}-1)}\equiv  \frac{\partial^{n-1}}{\partial X_{2a',2b'} \partial X_{2a_1,2b_1-1} \ldots \partial X_{2a_{n-2},2b_{n-2}-1}}.
\end{equation}
It is natural to fix $2a'$ and $2b'$ to be adjacent even integers~\footnote{In fact, using the relation between scaffolding variables and the usual Lorentz products~\cite{Arkani-Hamed:2023jry}, it is straightforward to show that $\partial_{X_{2a',2b'}}\sim \partial_{\epsilon_{a'} \cdot \epsilon_{b'}}$. }; without loss of generality, we choose $(2a', 2b') = (2, 2n)$. It is straightforward to see from~\eqref{eq:YM3pt} that for 3-point we have $\mathcal{D}_{(2,6)(1,4) }\mathcal{A}_3^{\rm YM}=1$. Experimentally, for $n=4$ we find~\footnote{The overall sign of the YM amplitude is fixed so that the ray-like diagram, where all points are connected to vertex-1, obtained from the general rules (described in the next subsection) corresponds to a positive sign. Consequently, all other diagrams will also carry a positive sign.}: 
\begin{equation}
\label{4pt}
    \mathcal{D}_{(2,8)(1,4)(1,6)} \mathcal{A}_4^{\rm YM}=\frac{1}{X_{1,5}}, \quad  \mathcal{D}_{(2,8)(1,4)(3,6)} \mathcal{A}_4^{\rm YM}=\frac{1}{X_{3,7}}\,,
\end{equation}
and similarly:
\begin{equation}
    \mathcal{D}_{(2,8)(4,7)(1,6)} \mathcal{A}_4^{\rm YM}=\frac{1}{X_{3,7}}, \quad  \mathcal{D}_{(2,8)(4,7)(3,6)} \mathcal{A}_4^{\rm YM}=-\frac{1}{X_{3,7}}\,.
\end{equation}


For $n=5$, we have:
\begin{equation}
    \mathcal{D}_{(2,10)(1,4)(1,6)(1,8)} \mathcal{A}_5^{\rm YM}=\frac{1}{X_{1,5}X_{1,7}}, \quad  \mathcal{D}_{(2,10)(1,4)(3,6)(3,8)} \mathcal{A}_5^{\rm YM}=\frac{1}{X_{3,7}X_{3,9}}\,.
\end{equation}
Clearly these are exactly the derivatives that produce the desired single cubic scalar diagrams. Of course there are operators that resulting in sum of cubic diagrams or $0$, for examples:
\begin{equation}
    \mathcal{D}_{(2,10)(1,4)(6,9)(5,8)} \mathcal{A}_5^{\rm YM}=-\frac{1}{X_{1,5} X_{5,9}}-\frac{1}{X_{3,9} X_{5,9}}, \quad  \mathcal{D}_{(2,10)(1,4)(1,6)(3,8)} \mathcal{A}_5^{\rm YM}=0\,.
\end{equation}
While it would be interesting to study the results of all possible inequivalent derivatives, for our purposes, it is particularly useful to first apply $\partial/{\partial X_{1,4}}$ (or $\partial/\partial X_{1,n-2}$) in addition to $\partial/\partial X_{2,2n}$. This choice ensures that the remaining $X_{e,o}$ derivatives correspond one-to-one with propagators $X_{o',o}$ (with $o'=e\pm1$) in the resulting diagram, according to our rules.  In the following subsections, we present the general rules and provide a proof for certain cases based on factorization.

\subsection{General rules and examples} 

For a given cubic diagram, we denote its propagators $X_{i,j}$ by $(i,j)$, where $1 \leqslant i < j \leqslant 2n-1$ and both $i$ and $j$ are odd integers. In this way, the diagram is represented by a sequence of diagonals $(i_1, j_1), \ldots, (i_{n-3}, j_{n-3})$, with each $(i,j)$ corresponding to a diagonal connecting vertices $i$ and $j$.

On the other hand, we denote the differential operators $\partial / \partial X_{a,b}$ by $[a,b]$, where $1 \leqslant a < b \leqslant 2n$. Thus the sequence of operators is written as $[a_1, b_1], \ldots, [a_{n-3}, b_{n-3}]$, with $[2,2n]$ and $[1,4]$ omitted for brevity. Note that since we always perform these two derivatives first, the subsequent steps will never involve operators like $[2,a]$, $[4,b]$, or $[2n,c]$. In particular, vertices 1 and 3 play a special role in the construction.

\begin{figure}[tbp]
    \centering
        \includegraphics[width=.7\textwidth]{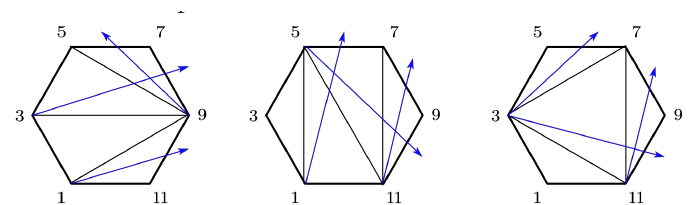}
        
     \caption{6-point $\phi ^3$ diagrams and corresponding differential operators for ray-like, zigzag and cyclic triangulations.\label{6-point}}
\end{figure}

By definition, they satisfy the symmetries $(i,j) = (j,i)$ and $[a,b] = [b,a]$. However, in certain cases, propagators may intentionally be written as $(j,i)$ with $i < j$ in order to apply specific rules. The correspondence between differential operators and single cubic diagrams is summarized below:
\begin{enumerate}
    \item For a given single cubic diagrams, reorganize the sequence of propagators \\$( i_1,j_1 ) ( i_2,j_2 ) ...( i_{n-3},j_{n-3} ) $ until it satisfies: if a propagator $(i_t,j_t)$ does not contain vertex-1 or 3 and $i_t\ne i_{t-1}$, then $i_t=j_{t-1} (2\leqslant t\leqslant n-3)$. In other words, the sequence should be made as consecutively connected as possible. Additionally, the sequence must start from vertex-1 or 3, since the subregion $\{1,2,3,2n\}$ must be triangulated, implying that at least one diagonal must be connected to vertex-1 or 3. For some cases where cycles appear (note that only cycles containing vertex-1 or 3 are considered), if vertex-1 is in the cycle, then the cycle should be written in a clockwise order in the sequence, otherwise the cycle should be written in a counterclockwise order. For examples, the 6-point zigzag and cycle diagrams (see figure~\ref{6-point}) $1/X_{1,5} X_{5,11} X_{7,11},$ $1/X_{3,7}X_{3,11}X_{7,11}$ is now labeled by $(1,5)(5,11)(11,7)$ and $(3,11)(11,7)(7,3)$, respectively.
    \item Next, we assign an operator to each propagator connected to vertex-1 or 3, according to the following rule:
    \begin{equation}
    \label{rule2}
        \begin{aligned}
        (1,j) \longleftrightarrow  [1, j+1], \\
        (3,j) \longleftrightarrow  [3, j-1], \\
        (i,1) \longleftrightarrow  [i+1, 1], \\
        (i,3) \longleftrightarrow  [i-1, 3].
    \end{aligned}
    \end{equation}
    For the 6-point zigzag diagram we mentioned above, we fix one derivative $(1,5) \leftrightarrow  [1, 6]$; and for the cycle we fix two $(3,11) \leftrightarrow  [3, 10]$, $(11,3) \leftrightarrow  [10, 3]$.
    \item The remaining differential operators are determined recursively as follows:
    \begin{equation}
    \label{rule3.1}
        \begin{array}{c}
	\cdots( 1,j ) ( i',j' ) \cdots\longleftrightarrow \cdots[ 1,j+1 ] [i',j'-1 ]\cdots\\
	\cdots( i,1 ) ( i',j' ) \cdots\longleftrightarrow \cdots[ i+1,1 ] [ i',j'-1 ] \cdots\\
\end{array} \quad ( j'\ne 1,3, i'\ne 1 ),
\end{equation}
\begin{equation}
 \label{rule3.2}
    \begin{array}{c}
	\cdots( 3,j ) ( i',j' ) \cdots\longleftrightarrow \cdots[ 3,j-1 ] [ i',j'+1 ] \cdots\\
	\cdots( i,3 ) ( i',j' ) \cdots\longleftrightarrow \cdots[ i-1,3 ] [ i',j'+1 ] \cdots\\
\end{array} \quad ( j'\ne 1,3, i'\ne 3 ),
\end{equation}
\begin{equation}
 \label{rule3.3}
        \cdots( i,j ) ( i',j' ) \cdots\longleftrightarrow \begin{cases}
	\cdots[ i,j\pm 1 ] [ i',j\pm 1 ] \cdots& \mathrm{if}\ i=i',\\
	\cdots[ i,j\pm 1 ] [ i',j\mp 1 ] \cdots&  \mathrm{otherwise}.\\
\end{cases}
\end{equation}
Moving back to the previous examples, for zigzag we have $(1,5)(5,11) \leftrightarrow  [1, 6][5,10]$ and then $(1,5)(5,11)(11,7) \leftrightarrow  [1, 6][5,10][11,8]$; similarly for cycle we conclude $(3,11)$ $(11,7)(7,3) \leftrightarrow  [3, 10][11,8][6,3]$.
    \item If there are more than one diagonal emanate from vertex- $i, (i\ne 1,3)$, first consider the diagonal that is closest to $(1,i)$ or $(3,i)$, If it is closer to $(1,i)$, then follow the rule that comes after $(1,i)$ (eq.\eqref{rule3.1}), else follow the rule that comes after $(3,i)$ (eq.\eqref{rule3.2}). Then the remaining correspondence can be determined by other rules.

\end{enumerate}

To summarize, we first write down the propagator sequence according to Rule 1, then determine the corresponding differential operators according to other rules (among these, Rule 2 has the highest priority). Next, we provide some examples to help the readers understand better (we use blue arrows to represent the position of differential operators in the $n$-polygon and omit $[ 2,2n ] ,[ 1,4 ] $).

\begin{figure}[tbp]
    \centering
    \includegraphics[width=.5\textwidth]{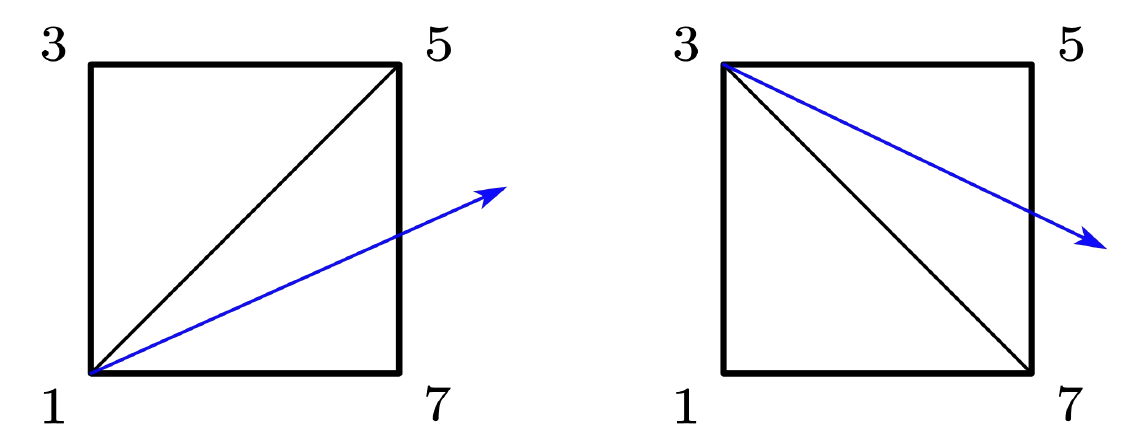}
    
     \caption{4-point $\phi ^3$ diagrams and corresponding differential operators. The propagators are represented by black diagonals, while differential operators are indicated by blue arrows. For example, the operator $[1,6]$ is depicted as an arrow starting from vertex-1 and pointing between vertices 5 and 7. \label{4-point}}
\end{figure}

\begin{example}
(4-point) In figure~\ref{4-point}, for the scalar diagram $1/X_{1,5}$, we first write the propagator as $(1,5)$, then according to \eqref{rule2}, the corresponding differential operators is $[1,6]$. For $1/X_{3,7}$, the propagator is $(3,7)$ and the corresponding differential operator is $[3,6]$, which are exactly eq.\eqref{4pt}.
\end{example}

\begin{figure}[tbp]
\centering
\includegraphics[width=.25\textwidth]{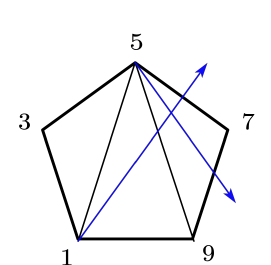}

\caption{5-point $\phi ^3$ diagram and corresponding differential operator.\label{5-point}}
\end{figure}

\begin{example}
(5-point) In figure~\ref{5-point}, for the scalar diagram $1/X_{1,5}X_{5,9}$, we first write the propagator sequence as $(1,5)(5,9)$ (according to Rule 1, we require the sequence must be connected as closely as possible front and back), then according to \eqref{rule2} and \eqref{rule3.1}, the corresponding differential operator is $[1,6][5,8]$. 
\end{example}

\begin{example}
(6-point ray-like) In figure~\ref{6-point}, for the scalar diagram $1/X_{1,9}X_{3,9}X_{5,9}$, we first write the propagator sequence as $(1,9)(3,9)(9,5)$, then according to \eqref{rule2} and \eqref{rule3.2}, the corresponding differential operator is $[1,10][3,8][9,6]$. 
\end{example}

\begin{example}
(6-point zigzag) In figure~\ref{6-point}, for the scalar diagram $1/X_{1,5}X_{5,11}X_{7,11}$, we first write the propagator sequence as $(1,5)(5,11)(11,7)$, then according to \eqref{rule2}, \eqref{rule3.1} and \eqref{rule3.3}, the corresponding differential operator is $[1,6][5,10][11,8]$. 
\end{example}

\begin{example}
(6-point cycle) In figure~\ref{6-point}, for the scalar diagram $1/X_{3,7}X_{3,11}X_{7,11}$, according to Rule 1, the propagators now have a cycle structure (3-7-11-3), since vertex-3 is in this cycle, so we should write the propagator sequence in a counterclockwise order: $(3,11)(11,7)(7,3)$, then according to \eqref{rule2} and \eqref{rule3.2}, the corresponding differential operator is $[3,10][11,8][6,3]$. 
\end{example}

\begin{figure}[tbp]
    \centering
        \includegraphics[width=.5\textwidth]{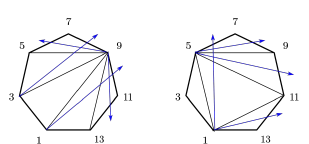}
        
    \caption{7-point $\phi ^3$ diagrams and corresponding differential operators for ray-like and cyclic triangulations.\label{7-point}}
\end{figure}

\begin{example}
(7-point ray-like) In figure~\ref{7-point}, for the scalar diagram $1/X_{1,9}X_{3,9}X_{5,9}X_{9,13}$, note that $(5,9)$ is closest to $(3,9)$ and $(9,13)$ is closest to $(1,9)$, so according to Rule 4 we first write the propagator sequence as $(1,9)(9,13)(3,9)(9,5)$, then according to \eqref{rule2}, \eqref{rule3.1} and  \eqref{rule3.2}, the corresponding differential operator is $[1,10][9,12][3,8][9,6]$. 
\end{example}

\begin{example}
(7-point cycle) In figure~\ref{7-point}, for the scalar diagram $1/X_{1,5}X_{1,11}X_{5,9}X_{5,11}$, according to Rule 1, the propagators now have a cycle structure (1-5-11-1), since vertex-1 is in this cycle, so we should write the cycle in a clockwise order: $(1,5)(5,9)(5,11)(11,1)$, then according to \eqref{rule2}, \eqref{rule3.1} and \eqref{rule3.3}, the corresponding differential operator is $[1,6][5,8][5,10][12,1]$.  
\end{example}

\begin{figure}[tbp]
    \centering
        \includegraphics[width=.5\textwidth]{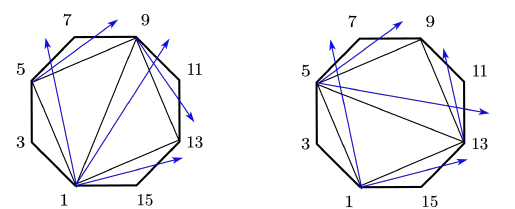}

    \caption{8-point $\phi ^3$ diagrams and corresponding differential operators.\label{8-point}}
\end{figure}

\begin{example}
(8-point cycle) In figure~\ref{8-point}, for the scalar diagram $1/X_{1,5}X_{1,9}X_{1,13}X_{5,9}X_{9,13}$, according to Rule 1, the propagators now have two cycles (1-5-9-1 and 1-9-13-1), since vertex-1 is in this cycle, so we should write the cycles in a clockwise order: $(1,5)(5,9)(1,9)(9,13)(13,1)$, then according to \eqref{rule2} and \eqref{rule3.1}, the corresponding differential operator is $[1,6]$$[5,8]$$[1,10]$$[9,12]$$[14,1]$.  
And for the scalar diagram $1/X_{1,5}X_{5,13}X_{1,13}X_{5,9}X_{9,13}$, according to Rule 1, the propagators now seem to have two cycles (1-5-13-1 and 5-9-13-5), but note that the second cycle does not contain vertex-1 or 3, so actually there is only one cycle. Since vertex-1 is in this cycle, so we should write the cycle in a clockwise order: $(1,5)(5,9)(5,13)(13,9)(13,1)$, then according to \eqref{rule2}, \eqref{rule3.1} and \eqref{rule3.3}, the corresponding differential operator is $[1,6][5,8][5,12][13,10][14,1]$.  
\end{example}

Let us summarize the evident pattern for certain special topologies at general $n$.
\begin{example}
($n$-point ray-like originating at vertices 1 and 3) For the scalar diagram
\begin{equation}
    \frac{1}{X_{1,5}X_{1,7}X_{1,9}\cdots X_{1,2n-3}},
\end{equation}
the propagator sequence is $(1,5)(1,7)(1,9)\cdots(1,2n-3)$ and the corresponding differential operator is given by $[1,6][1,8][1,10]\cdots[1,2n-2]$.
\newline
Analogously, for the scalar diagram
\begin{equation}
    \frac{1}{X_{3,7}X_{3,9}X_{3,11}\cdots X_{3,2n-1}},
\end{equation}
the propagator sequence is $(3,7)(3,9)(3,11)\cdots(3,2n-1)$ and the corresponding differential operator is $[3,6][3,8][3,10]\cdots[3,2n-2]$.

We will give a thorough proof of these two special cases in the next subsection.
\end{example}

\begin{example}
($n$-point ray-like originating at vertices 5 and $2n-1$) For the scalar diagram
\begin{equation}
    \frac{1}{X_{5,1}X_{5,9}X_{5,11}\cdots X_{5,2n-1}},
\end{equation}
the propagator sequence is $(1,5)(5,2n-1)(5,2n-3)\cdots(5,9)$ and the corresponding differential operator is $[1,6][5,2n-2][5,2n-4]\cdots[5,8]$.

For the scalar diagram
\begin{equation}
    \frac{1}{X_{2n-1,3}X_{2n-1,5}X_{2n-1,7}\cdots X_{2n-1,2n-5}},
\end{equation}
the propagator sequence is $(3,2n-1)(2n-1,5)(2n-1,7)\cdots(2n-1,2n-5)$ and the corresponding differential operator is $[3,2n-2][2n-1,6][2n-1,8]\cdots[2n-1,2n-4]$.
\end{example}

\begin{example}
($n$-point ray-like originating at vertices 7 and $2n-3$) For the scalar diagram
\begin{equation}
    \frac{1}{X_{7,1}X_{7,3}X_{7,11}\cdots X_{7,2n-1}},
\end{equation}
the propagator sequence is $(3,7)(1,7)(7,2n-1)(7,2n-3)\cdots(7,11)$ and the corresponding differential operator is $[3,6][1,8][7,2n-2][7,2n-4]\cdots[7,10]$.

For the scalar diagram
\begin{equation}
    \frac{1}{X_{2n-3,1}X_{2n-3,3}X_{2n-3,5}\cdots X_{2n-3,2n-7}},
\end{equation}
the propagator sequence is $(1,2n-3)(3,2n-3)(2n-3,5)\cdots(2n-3,2n-7)$ and the corresponding differential operator is $[1,2n-2][3,2n-4][2n-3,6]\cdots[2n-3,2n-6]$.
\end{example}
\begin{example}
($n$-point ray-like) For the scalar diagram ($i\ne 1,3,5,7,2n-3,2n-1$)
\begin{equation}
    \frac{1}{X_{i,1}X_{i,3}X_{i,5}\cdots X_{i,i-4}X_{i,i+4}\cdots X_{i,2n-1}},
\end{equation}
according to Rule 4, we should write the propagator that is closest to $(1,i)$ or $(3,i)$, so the propagator sequence is $(1,i)(i,2n-1)(i,2n-3)\cdots(i,i+4)(3,i)(i,5)(i,7)\cdots(i,i-4)$ and the corresponding differential operator is $[1,i+1][i,2n-2][i,2n-4]\cdots[i,i+3][3,i-1][i,6][i,8]\cdots[i,i-3]$.
\end{example}

\begin{example}
($n$-point zigzag starting from vertex-1) For the scalar diagram
\begin{equation}
    \frac{1}{X_{1,5}X_{5,2n-1}X_{2n-1,7}X_{7,2n-3}\cdots},
\end{equation}
the propagator sequence is $(1,5)(5,2n-1)(2n-1,7)(7,2n-3)\cdots$ and the corresponding differential operator is $[1,6][5,2n-2][2n-1,8][7,2n-4]\cdots$.
\newline
For the scalar diagram
\begin{equation}
    \frac{1}{X_{1,2n-3}X_{2n-3,3}X_{3,2n-5}X_{2n-5,5}X_{5,2n-7}\cdots},
\end{equation}
the propagator sequence is $(1,2n-3)(3,2n-3)(3,2n-5)(2n-5,5)(5,2n-7)\cdots$ and the corresponding differential operator is $[1,2n-2][3,2n-4][3,2n-6][2n-5,6][5,2n-8]\cdots$.
\end{example}

\begin{example}
($n$-point one-flip from ray-like) For the scalar diagram
\begin{equation}
    \frac{1}{X_{1,5}X_{1,7}X_{1,9}\cdots X_{2n-7,2n-3}X_{1,2n-3}},
\end{equation}
it is easy to see that in this case we just flip the $(1,2n-5)$ in the 1-ray-like to $(2n-7,2n-3)$. Now the propagator sequence is $(1,5)(1,7)(1,9)\cdots(1,2n-7)(2n-7,2n-3)(1,2n-3)$ (note that after flipping, there is a cycle which has vertex-1 so we should write this cycle in a clockwise order) and the corresponding differential operator is $[1,6][1,8][1,10]\cdots[1,2n-6][2n-7,2n-4][1,2n-2]$.
\end{example}


\subsection{Proof of the rules by factorization}
We have presented the rules for constructing operators that extract single cubic diagrams from YM amplitudes. We now provide a proof of these rules, focusing on certain special cases at general multiplicity. The proof is divided into two main steps, based on factorization. First, for a given diagram, we show that the differential operator acting on the corresponding $(n-3)$ consecutive factorizations (residues) yields 1. Second, we demonstrate that the same differential operator vanishes when acting on other consecutive factorizations associated with different diagrams. Each step is established recursively.

Let us begin by proving the validity of the operator associated with the ray-like diagram in which all vertices are connected to vertex-$1$.

\begin{figure}[tbp]
    \centering
        \includegraphics[width=.6\textwidth]{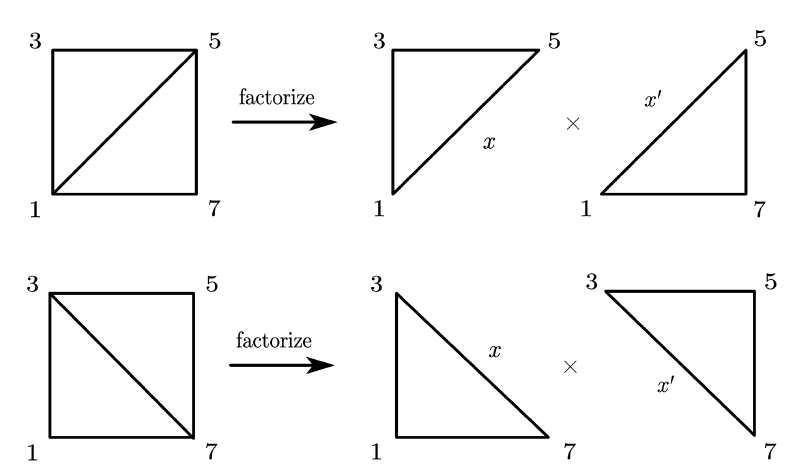}
    
    \caption{4-point factorizations.\label{4pt_fac}}
\end{figure}

\begin{figure}[tbp]
    \centering
        \includegraphics[width=.6\textwidth]{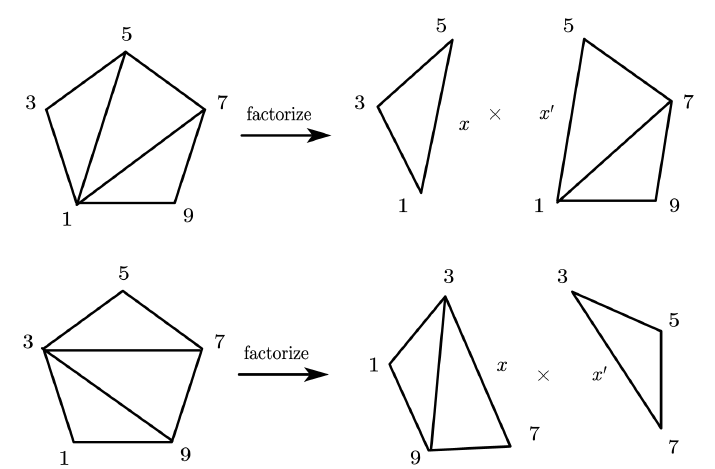}
    
    \caption{5-point factorizations.\label{5pt_fac}}
\end{figure}

\begin{figure}[tbp]
    \centering
        \includegraphics[width=.6\textwidth]{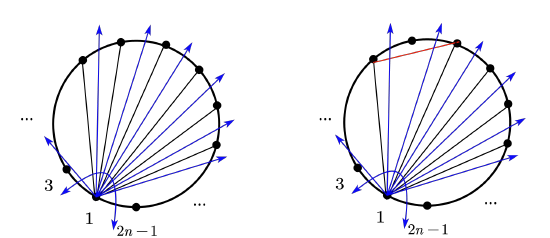}

    \caption{The number of vertex-1 ray-like intersections is $n-3$ (left). In other cases, for example after one flip (denoted by the red line) from vertex-1 ray-like, the number of intersections is more than $n-3$  (right).\label{1ray_cross}}
\end{figure}

\begin{figure}[tbp]
    \centering
        \includegraphics[width=.6\textwidth]{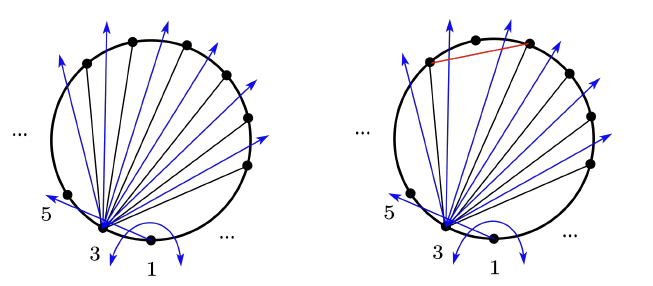}

    \caption{The number of vertex-3 ray-like intersections is $n-3$ (left). In other cases, for example after one flip (denoted by the red line) from vertex-3 ray-like, the number of intersections is more than $n-3$  (right).\label{3ray_cross}}
\end{figure}

\textbf{Step I.} We aim to show:
\begin{equation}
    \mathcal{D}_{(2,2n),(1,4),(1,6),\ldots,(1,2n-2)} {\rm Res}_{X_{1,5}=X_{1,7}=\ldots=X_{1,2n-3}=0} {\cal A}_n^{\rm YM}=1 \,.
\end{equation}
To begin with, consider the case $n = 4$, for which we have:
\begin{align} \label{eq:4ptfac}
    &\mathcal{D}_{(2,8),(1,4),(1,6)} {\rm Res}_{X_{1,5}=0} \mathcal{A}_4(1,2,\ldots,8)
    \,,  \\
    =&\mathcal{D}_{(2,8),(1,4),(1,6)} \sum_{j,J} (X_{j,J}-X_{1,j}-X_{5,J}) \frac{\partial }{\partial X_{x,j}} \mathcal{A}_L(1,2,3,4,5,x) \, \frac{\partial }{\partial X_{x',J}} \mathcal{A}_R(5,6,7,8,1,x') \,, \nonumber
\end{align}
where we have used~\eqref{eq:fac1}, and the left and right 3-point subamplitudes are labeled by $(1,2,3,4,5,x)$ and $(5,6,7,8,1,x')$, respectively (figure~\ref{4pt_fac}). In this case, the derivative $\partial / \partial X_{2,8}$ must act on the prefactor, leading to:
\begin{equation}
    \mathcal{D}_{(1,4),(1,6)} \frac{\partial }{\partial X_{x,2}} \mathcal{A}_L(1,2,3,4,5,x) \, \frac{\partial }{\partial X_{x',8}} \mathcal{A}_R(5,6,7,8,1,x')=1 \,.
\end{equation}
In the final step, we have used the explicit expression~\eqref{eq:YM3pt} with the appropriate relabeling. Moving on to the case $n = 5$ (figure~\ref{5pt_fac}), we find:
\begin{align}
   & \mathcal{D}_{(2,10),(1,4),(1,6),(1,8)} \sum_{j,J} (X_{j,J}-X_{1,j}-X_{5,J}) \frac{\partial }{\partial X_{x,j}} \mathcal{A}_L(1,2,3,4,5,x) \\& \times {\rm Res}_{X_{1,7}=0}\frac{\partial }{\partial X_{x',J}} \mathcal{A}_R(5,6,7,8,9,10,1,x') \\
    =& \mathcal{D}_{(1,4),(1,6),(1,8)} \frac{\partial }{\partial X_{x,2}} \mathcal{A}_L(1,2,3,4,5,x) \times {\rm Res}_{X_{1,7}=0}\frac{\partial }{\partial X_{x',10}} \nonumber\mathcal{A}_R(5,6,7,8,9,10,1,x') \\
    =& \mathcal{D}_{(x',10),(1,6),(1,8)} {\rm Res}_{X_{1,7}=0} \mathcal{A}(1,x',5,6,7,8,9,10)=1
    \,.  \nonumber
\end{align}
In the second line, we have used the fact that $\partial/\partial X_{2,10}$ must act on the prefactor. In the final line, we have rewritten the result (after a cyclic relabeling of external legs) to match exactly with~\eqref{eq:4ptfac}. This argument can be applied recursively to the general $n$-point case:
\begin{align}
    &\mathcal{D}_{(2,2n),(1,4),(1,6),\ldots,(1,2n-2)} {\rm Res}_{X_{1,5}=X_{1,7}=\ldots=X_{1,2n-3}=0} {\cal A}_n^{\rm YM}(1,2,\ldots,2n)\\  \nonumber
    =& \mathcal{D}_{(x',2n),(1,6),\ldots,(1,2n-2)} {\rm Res}_{X_{1,7}=X_{1,9}=\ldots=X_{1,2n-3}=0} {\cal A}_{n-1}^{\rm YM}(1,x',5,6,\ldots,2n)\\ \nonumber
    =&\ldots=1 \,.
\end{align}

\textbf{Step II.} We now demonstrate:
\begin{equation}
    \mathcal{D}_{(2,2n),(1,4),(1,6),\ldots,(1,2n-2)} {\rm Res}_{X_{i_1,j_1}=X_{i_2,j_2}=\ldots=X_{i_{n-1},j_{n-1}}=0} {\cal A}_n^{\rm YM}=0 \, ,
\end{equation}
for any sequence $(i_1,j_1)(i_2,j_2)\ldots(i_{n-1},j_{n-1}) \ne (1,5)(1,7)\ldots(1,2n-3)$. In the previous step, we saw that if a derivative involves a variable $X_{a,b}$ where $a$ and $b$ appear on opposite sides of a factorization channel, then that derivative must act on the prefactor in~\eqref{eq:fac1}.

Now, suppose we draw the derivatives $[2,2n], [1,4], [1,6], \ldots, [1,2n-2]$ as blue lines, and the propagators $(i_1,j_1)(i_2,j_2)\ldots(i_{n-1},j_{n-1})$ as black lines. Then, for the derivative not to vanish, the number of intersections between blue and black lines must be at most $n-3$.In figure~\ref{1ray_cross}, it is direct to find that there is only one such configuration with exactly $n-3$ intersections, namely, when $(i_1,j_1)(i_2,j_2)\ldots(i_{n-1},j_{n-1}) = (1,5)(1,7)\ldots(1,2n-3)$.

This completes the proof.

Following similar steps, we now prove the validity of the operator associated with the ray-like diagram in which all vertices are connected to vertex-3.

\textbf{Step I.} We aim to show:
\begin{equation}
    \mathcal{D}_{(2,2n),(1,4),(3,6),(3,8),\ldots,(3,2n-2)} {\rm Res}_{X_{3,7}=X_{3,9}=\ldots=X_{3,2n-1}=0} {\cal A}_n^{\rm YM}=1 \,.
\end{equation}
For the case $n = 4$, we have:
\begin{align} \label{eq:4ptfac'}
    &\mathcal{D}_{(2,8),(1,4),(3,6)} {\rm Res}_{X_{3,7}=0} \mathcal{A}_4(1,2,\ldots,8)
    \,,  \\
    =&\mathcal{D}_{(2,8),(1,4),(3,6)} \sum_{j,J} (X_{j,J}-X_{3,j}-X_{7,J}) \frac{\partial }{\partial X_{x,j}} \mathcal{A}_L(3,x,7,8,1,2) \, \frac{\partial }{\partial X_{x',J}} \mathcal{A}_R(3,4,5,6,7,x') \,, \nonumber
\end{align}
now the left and right 3-point subamplitudes are labeled by $(3,x,7,8,1,2)$ and $(3,4,5,6,7,x')$, respectively (figure~\ref{4pt_fac}). In this case, the derivative $\partial / \partial X_{1,4}$ must act on the prefactor, leading to:
\begin{equation}
    \mathcal{D}_{(2,8),(3,6)} \frac{\partial }{\partial X_{x,1}} \mathcal{A}_L(3,x,7,8,1,2) \, \frac{\partial }{\partial X_{x',4}} \mathcal{A}_R(3,4,5,6,7,x')=1 \,.
\end{equation}
In the final step, we have used the explicit expression~\eqref{eq:YM3pt} with the appropriate relabeling. For the case $n = 5$ (figure~\ref{5pt_fac}), we have:
\begin{align}
   & \mathcal{D}_{(2,10),(1,4),(3,6),(3,8)} \sum_{j,J} (X_{j,J}-X_{3,j}-X_{7,J}) \frac{\partial }{\partial X_{x',j}} \mathcal{A}_R(3,4,5,6,7,x') \\& \times {\rm Res}_{X_{3,9}=0}\frac{\partial }{\partial X_{x,J}} \mathcal{A}_L(3,x,7,8,9,10,1,2) \\
    =& \mathcal{D}_{(2,10),(3,6),(3,8)} \frac{\partial }{\partial X_{x',4}} \mathcal{A}_R(3,4,5,6,7,x') \times {\rm Res}_{X_{3,9}=0}\frac{\partial }{\partial X_{x,1}} \nonumber\mathcal{A}_L(3,x,7,8,9,10,1,2) \\
    =& \mathcal{D}_{(x,1),(2,10),(3,8)} {\rm Res}_{X_{3,9}=0} \mathcal{A}(1,2,3,x,7,8,9,10)=1
    \,.  \nonumber
\end{align}
In the second line, we have used the fact that $\partial/\partial X_{1,4}$ must act on the prefactor. In the final line, we have rewritten the result (after a cyclic relabeling of external legs) to match exactly with~\eqref{eq:4ptfac'}. This argument can be applied recursively to the general $n$-point case:
\begin{align}
    &\mathcal{D}_{(2,2n),(1,4),(3,6),(3,8),\ldots,(3,2n-2)} {\rm Res}_{X_{3,7}=X_{3,9}=\ldots=X_{3,2n-1}=0} {\cal A}_n^{\rm YM}(1,2,\ldots,2n)\\ \nonumber
    =& \mathcal{D}_{(x,1),(2,2n),(3,8)\ldots,(3,2n-2)} {\rm Res}_{X_{3,9}=X_{3,11}=\ldots=X_{3,2n-1}=0} {\cal A}_{n-1}^{\rm YM}(1,2,3,x,7,8,\ldots,2n)\\ \nonumber
    =&\ldots=1 \,.
\end{align}

\textbf{Step II.} We now demonstrate:
\begin{equation}
    \mathcal{D}_{(2,2n),(1,4),(3,6),(3,8)\ldots,(3,2n-2)} {\rm Res}_{X_{i_1,j_1}=X_{i_2,j_2}=\ldots=X_{i_{n-1},j_{n-1}}=0} {\cal A}_n^{\rm YM}=0 \, ,
\end{equation}
for any sequence $(i_1,j_1)(i_2,j_2)\ldots(i_{n-1},j_{n-1}) \ne (3,7)(3,9)\ldots(3,2n-1)$. In figure~\ref{3ray_cross}, similar to the vertex-1 ray-like case, for the derivative not to vanish, the number of intersections between blue and black lines must be at most $n-3$. It is direct to find that there is only one such configuration with exactly $n-3$ intersections, namely, when $(i_1,j_1)(i_2,j_2)\ldots(i_{n-1},j_{n-1}) = (3,7)(3,9)\ldots(3,2n-1)$.

This completes the proof.

\subsection{Remarks on different approaches to subtracting scalar amplitudes from gluon amplitudes}

As mentioned earlier, the previous literature~\cite{Cheung:2017ems,Backus:2025njt} has introduced different approaches to subtracting scalar amplitudes from gluon amplitudes. In~\cite{Cheung:2017ems}, the differential operators are constructed in terms of conventional Lorentz products, while in~\cite{Backus:2025njt} the operators are also formulated using scalar-scaffolded variables. In this subsection, we comment on the relations and differences between these approaches. Let us illustrate this with a 4-point example by first comparing with the operator constructed in~\cite{Backus:2025njt}, which computes the scalar amplitude with an additional $X_{e,e}$ factor, {\it e.g.} $X_{2,8} \mathcal{A}_4^{\phi^3}(1,2,3,4)$ via a sum of $(n\!-\!2)$-fold operators which generally takes the form$\prod_{e} \sum_{i\neq e,e\pm1} \partial /\partial X_{e,i}$, acting on the YM amplitude:
\begin{equation} \label{eq: 2505 operator}
\begin{aligned}
  X_{2,8} \mathcal{A}_4^{\phi^3}(1,2,3,4)=\sum_{i=1,2,7,8}  \frac{\partial}{\partial X_{4,i}} \sum_{j=1,2,3,8}  \frac{\partial}{\partial X_{6,j}}  \mathcal{A}_4^{\rm YM} =X_{2,8}\left(\frac{1}{X_{1,5}}+\frac{1}{X_{3,7}}\right) \,,
\end{aligned}
\end{equation}
where the tr$(\phi^3)$ amplitude, or equivalently, the bi-adjoint scalar amplitude with both orderings~\footnote{Note that in our notation an overall canonical ordering is implicit throughout.} taken to be the canonical ordering is given by the sum of two cubic diagrams. Meanwhile, in our approach, we construct each diagram individually, without any prefactor, using a single $(n\!-\!1)$-fold operator: 
\begin{align} \label{eq: 2512 operator1}
    &-\mathcal{A}_4^{\phi^3}(1,2,4,3)= \frac{\partial}{\partial X_{2,8}} \frac{\partial}{\partial X_{1,4}}   \frac{\partial}{\partial X_{1,6}}  \mathcal{A}_4^{\rm YM} =\frac{1}{X_{1,5}}\\
    \label{eq: 2512 operator2}
    &-\mathcal{A}_4^{\phi^3}(1,3,2,4)= \frac{\partial}{\partial X_{2,8}} \frac{\partial}{\partial X_{1,4}}   \frac{\partial}{\partial X_{3,6}}  \mathcal{A}_4^{\rm YM} =\frac{1}{X_{3,7}}\,,
\end{align}
where, for later convenience in comparing with the operators in~\cite{Cheung:2017ems}, we emphasize on the left-hand side of the first equality that each diagram can also be viewed as a bi-adjoint $\phi^3$ amplitude with the second ordering taken to be a specific ordering. 

There are two main differences between the above two approaches. The $(n\!-\!2)$-fold operators~\eqref{eq: 2505 operator} involve more terms (growing roughly factorially), but they do not require taking derivatives with respect to $X_{e,e}$. On the other hand, in the $(n\!-\!1)$-fold operator construction, each diagram is produced by a single operator, so only $\mathcal{C}_{n-2}$ terms are needed to obtain the tr$(\phi^3)$ amplitude. However, derivatives with respect to $X_{e,e}$ cannot be avoided in this approach; for example,
\begin{equation}
     \frac{\partial}{\partial X_{1,4}}   \frac{\partial}{\partial X_{1,6}}  \mathcal{A}_4^{\rm YM} \neq X_{2,8}\times \frac{1}{X_{1,5}}\,.
\end{equation}
Next, we consider the operators introduced in~\cite{Cheung:2017ems}, which we reviewed in Section~\ref{sec: trans operators}. These operators, constructed in terms of conventional Lorentz products, can be used to obtain bi-adjoint $\phi^3$ amplitudes with any chosen second ordering. Moreover, given the relations between the Lorentz products and the scalar-scaffolded variables~\eqref{eq: epXtrans}, it is straightforward to express these operators in terms of scalar-scaffolded variables using the chain rule:
\begin{equation}
\begin{aligned}
   \mathcal{A}_4^{\phi^3}(1,2,3,4)=&   \frac{\partial}{\partial \epsilon_1 \cdot \epsilon_4} \left(\frac{\partial}{\partial \epsilon_2 \cdot k_1}-\frac{\partial}{\partial \epsilon_2 \cdot k_4}\right) \left(\frac{\partial}{\partial \epsilon_3 \cdot k_2}-\frac{\partial}{\partial \epsilon_3 \cdot k_4}\right) \mathcal{A}_4^{\rm YM} \\
   \to &\frac{\partial}{\partial X_{2,8}} \sum_{i=1,2}  \frac{\partial}{\partial X_{4,i}} \sum_{j=1,2,3,4}  \frac{\partial}{\partial X_{6,j}}  \mathcal{A}_4^{\rm YM} \\ 
   =&\frac{1}{X_{1,5}}+\frac{1}{X_{3,7}} \,,
\end{aligned}
\end{equation}

\begin{equation}
\begin{aligned}
   \mathcal{A}_4^{\phi^3}(1,2,4,3)=&   \frac{\partial}{\partial \epsilon_1 \cdot \epsilon_3} \left(\frac{\partial}{\partial \epsilon_2 \cdot k_1}-\frac{\partial}{\partial \epsilon_2 \cdot k_3}\right) \left(\frac{\partial}{\partial \epsilon_4 \cdot k_2}-\frac{\partial}{\partial \epsilon_4 \cdot k_3}\right) \mathcal{A}_4^{\rm YM} \\
   \to &-\frac{\partial}{\partial X_{2,6}} \sum_{i=1,2,7,8}  \frac{\partial}{\partial X_{4,i}} \sum_{j=5,6}  \frac{\partial}{\partial X_{4,j}}  \mathcal{A}_4^{\rm YM} \\ 
   =&-\frac{1}{X_{1,5}} \,,
\end{aligned}
\end{equation}

\begin{equation}
\begin{aligned}
   \mathcal{A}_4^{\phi^3}(1,3,2,4)=&   \frac{\partial}{\partial \epsilon_1 \cdot \epsilon_4} \left(\frac{\partial}{\partial \epsilon_3 \cdot k_1}-\frac{\partial}{\partial \epsilon_3 \cdot k_4}\right) \left(\frac{\partial}{\partial \epsilon_2 \cdot k_3}-\frac{\partial}{\partial \epsilon_2 \cdot k_4}\right) \mathcal{A}_4^{\rm YM} \\
   \to &-\frac{\partial}{\partial X_{2,8}} \sum_{i=1,2}  \frac{\partial}{\partial X_{6,i}} \sum_{j=7,8}  \frac{\partial}{\partial X_{8,j}}  \mathcal{A}_4^{\rm YM} \\ 
   =&-\frac{1}{X_{3,7}} \,.
\end{aligned}
\end{equation}
As we have seen, although derivatives with respect to $X_{e,e}$ are unavoidable as in~\eqref{eq: 2512 operator1} and~\eqref{eq: 2512 operator2}, the operators above also involve more terms, as in~\eqref{eq: 2505 operator}.

In summary, the three approaches discussed above are distinct and each has its own features. To some extent, the operators in~\cite{Cheung:2017ems} are more flexible, since the second ordering in bi-adjoint $\phi^3$ can be specified arbitrarily. The operators~\eqref{eq: 2505 operator} have a clear string-theoretic origin, as discussed in~\cite{Backus:2025njt}, and possess the notable feature that they do not require derivatives with respect to $X_{e,e}$. On the other hand, the operators constructed in this paper establish a more direct connection to the structure of YM amplitudes, as each term is obtained from a single operator.

\section{Gluon Insertions and Minimal Basis}\label{sec: minimal basis}
We have introduced a systematic method for extracting single-scalar cubic diagrams through differential operators. A natural question is whether there exist analogous operators that yield amplitudes containing gluons.  In this section, we start by counting the number of independent ``mixed amplitudes'' generated by these operators. As we will see, planarity dictates that this number is given by $\mathcal{C}_{r-2}$, where $r$ denotes the number of external scalar legs.

To be concrete, let us consider the number of linearly independent mixed amplitudes with chosen unordered gluons denoted by:
\begin{equation}
    N_{n,r} = \dim \mathrm{Span}\{\mathcal{A}_n^{\mathrm{YM}+\phi^3}(\{\bar{\beta}\}|\beta)\},
\end{equation}
here $|\beta| = r$ denotes the number of scalars and $\{\bar{\beta}\}$ the complementary set of $\beta$, representing the gluons. Note that on the right hand side we have collected all possible permutations of $\beta$. It is clear that this number is independent of which particles are chosen as scalars, but only on their total number $r$\footnote{To provide simple examples, we only consider modding out the cyclic and reflection symmetries, the naive dimension is $\frac{(n-1)!}{2}$. One may further impose the Kleiss-Kuijf relation~\cite{Kleiss:1988ne,DelDuca:1999rs} to fix two legs, reducing the naive dimension to $(n-2)!$. However, the actual number is much smaller than either of these.}. As we will see later, the number of gluons $n-r$, or equivalently the total number of particles $n$, is also irrelevant, {\it i.e.} we have $N_{n,r} = N_r$.

To illustrate, let us consider the case $n=4$ with $r=2,3,4$. Obviously, we have $N_{4,2} = N_{4,3} = 1$ due to the cyclic and reflection symmetries of the colored scalars. Furthermore, in the pure scalar case, {\it i.e.}, $r=4$, the result is given by the number of 4-point ordered cubic diagrams: $N_{4,4} =2$. In fact, at any given multiplicity we have $N_{n,2}=N_{n,3}=1$ and $N_{n,n}=\mathcal{C}_{n-2}$.

\begin{figure}
    \centering
    \includegraphics[width=0.6\linewidth]{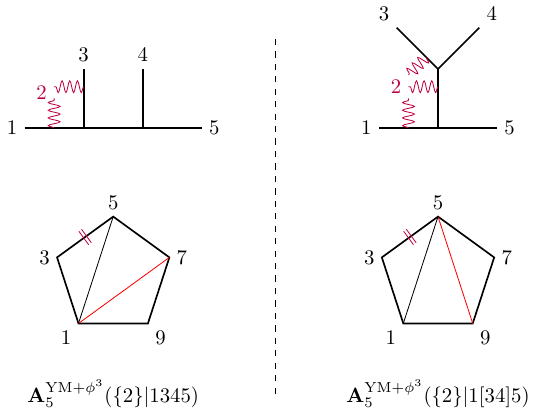}
    \caption{Single-gluon insertion at 5-point level, in which insertion in  $s$ and $t$-channels gives two independent amplitudes. In the pentagon, the red curves denote the propagators inherited from the 4-point skeleton diagrams, while the black ones separate out an effective scalar, yielding an equivalent lower-point pure scalar problem.}
    \label{fig:gluon_Feyman1}
\end{figure}

The first nontrivial case appears at $n=5$ with $r=4$. Without loss of generality, we choose the scalars to be $1,3,4,5$, leaving $2$ as the gluon. Taking into account the redundancies from cyclic and reflection symmetries, there are naively three equivalent orderings, namely
\begin{equation}
    \{\mathcal{A}_5^{\mathrm{YM}+\phi^3}(\{2\}|1, 3, 4, 5),\,
      \mathcal{A}_5^{\mathrm{YM}+\phi^3}(\{2\}|1, 3, 5, 4),\,
      \mathcal{A}_5^{\mathrm{YM}+\phi^3}(\{2\}|1, 4, 3, 5)\}\,.
\end{equation}
Given the advantage of scaffolding variables, which provide a unique representation, we write
\begin{equation*}
    \mathcal{A}_5^{\mathrm{YM}+\phi^3}(\{2\}|\beta)= n(\beta) \cdot \left\{\frac{X_{4,9}}{X_{3,9} X_{5,9}},\frac{X_{1,4}}{X_{1,5} X_{5,9}},\frac{X_{4,7}}{X_{3,7} X_{3,9}},\frac{X_{4,7}}{X_{1,7} X_{3,7}},\frac{X_{1,4}}{X_{1,5} X_{1,7}},\frac{1}{X_{5,9}},\frac{1}{X_{3,9}},\frac{1}{X_{1,7}}\right\} ,
\end{equation*}
where $n(\beta)$ is an 8-dimensional vector. Then we obtain the following matrix:
\begin{equation}
   \left(\begin{array}{c}
      n(1,3,4,5)  \\
     n(1,3,5,4)   \\
      n(1,4,3,5)  \\
   \end{array} \right)=\left(
\begin{array}{cccccccc}
 -1 & -1 & -1 & -1 & -1 & 1 & 1 & 1 \\
 0 & 0 & 0 & 1 & 1 & 0 & 0 & -1 \\
 1 & 1 & 1 & 0 & 0 & -1 & -1 & 0 \\
\end{array}
\right)\,,
\end{equation}
where we have $n(1,3,4,5)=-n(1,3,5,4)-n(1,4,3,5)$ and therefore the above matrix has rank 2. As a consequence, we find
\begin{equation}
N_{5,4} = 2.
\end{equation}
Moving on to $n=6$, there are two nontrivial cases with $r=4$ and $r=5$, respectively.
For $r=4$, one can similarly verify that the corresponding $3 \times 74$ matrix has rank 2. For $r=5$, as shown in Appendix~\ref{app: rank example}, the $12 \times 23$ matrix has rank 5, and thus
\begin{equation} 
N_{6,5} = \mathcal{C}_{3} = 5 \,.
\end{equation}
So far we have experimentally found
\begin{equation}  \label{eq: number_statement}
N_{r} \equiv N_{n,r} = \mathcal{C}_{r-2} \, .
\end{equation}
This is perhaps not surprising. Let us consider one step away from the pure scalar case, {\it i.e.}, with a single gluon insertion. For $n=5$ with scalar $1,3,4,5$, there are two possible scalar skeletons ($s$ and $t$ channels), and the amplitude is obtained by inserting one gluon into these skeletons in all possible ways that preserve the overall ordering. That said, there exists a natural basis for such amplitudes:
\begin{equation}
\begin{aligned}
    &\mathbf{A}_5^{\mathrm{YM}+\phi^3}(\{2\}|1345)= \mathcal{A}_5^{\mathrm{YM}+\phi^3}(\{2\}|1, 3, 5, 4),\quad \\
      & \mathbf{A}_5^{\mathrm{YM}+\phi^3}(\{2\}|1[34]5)=-\mathcal{A}_5^{\mathrm{YM}+\phi^3}(\{2\}|1, 4, 3, 5)\,, 
\end{aligned}
\end{equation}
Here, $\mathbf{A}_5^{\mathrm{YM}+\phi^3}(\{2\}|1345)$ and $\mathbf{A}_5^{\mathrm{YM}+\phi^3}(\{2\}|1[34]5)$ denote the five-point mixed amplitudes with gluon $2$ inserted into the scalar skeletons shown on the left and right of Figure~\ref{fig:gluon_Feyman1}, respectively. We use $1345$ to label the diagram with no subtree on the line $(1,n)$, and $1[34]5$ to label the diagram containing the subtree $[34]$ on the same line. Such subtree structures correspond one-to-one with commutators. For later convenience, we also represent these amplitudes using a triangulated 5-gon. In this diagram, the edge $(35)$ is marked with a purple double line to denote the gluon, while the black diagonal $(15)$ separates the region into to an effective pure scalar subdiagram $(1579)$. Furthermore, the red diagonals $(17)$ and $(59)$ are used to indicate the $s$- and $t$-channels, respectively~\footnote{The red and black diagonals \textbf{do not} correspond to the final propagators after the gluon insertion.}.

\begin{figure}
    \centering
    \includegraphics[width=0.8\linewidth]{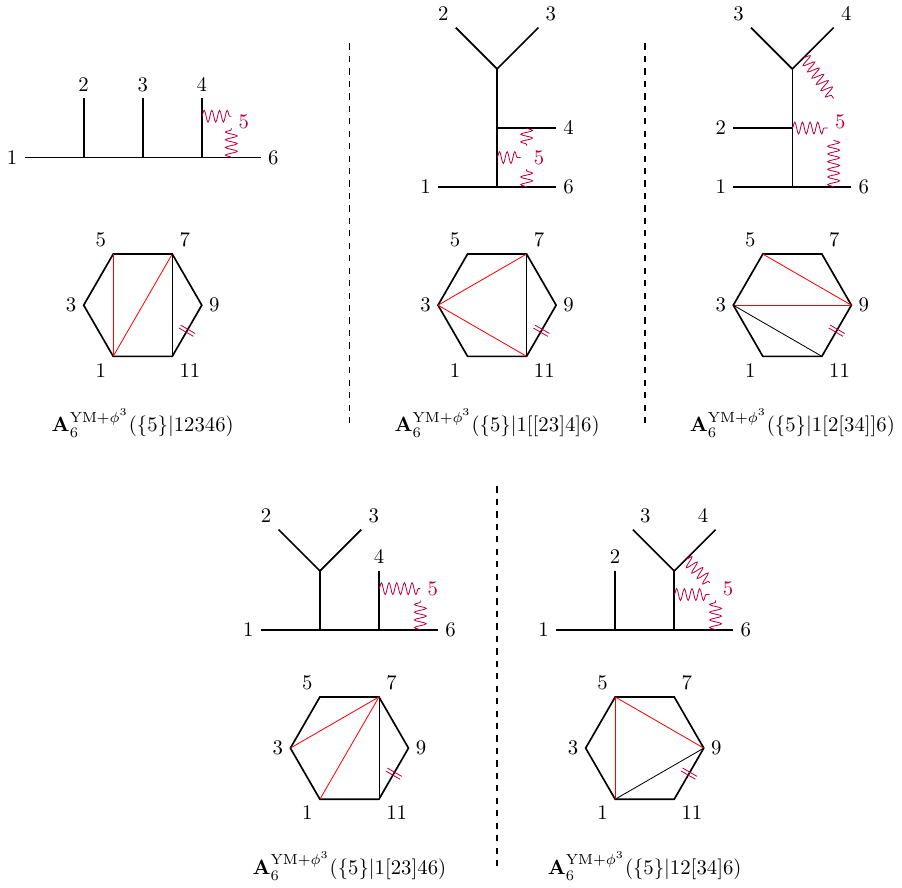}
    \caption{Single-gluon insertion for $n=6$, with scalars being 1,2,3,4,6 and gluon being 5.}
    \label{fig:gluon_Feyman2}
\end{figure}

Similarly, the linear space for $n=6$ with scalar $1,2,3,4,6$ is spanned by (see Figure~\ref{fig:gluon_Feyman2}):
\begin{equation}
\begin{aligned}
&\mathbf{A}_6^{\mathrm{YM}+\phi^3}(\{5\}|12346)\, ,  \mathbf{A}_6^{\mathrm{YM}+\phi^3}(\{5\}|1[[23]4]6) \, ,  \mathbf{A}_6^{\mathrm{YM}+\phi^3}(\{5\}|1[2[34]]6) \, , \\  &\mathbf{A}_6^{\mathrm{YM}+\phi^3}(\{5\}|1[23]46) \, ,  \mathbf{A}_6^{\mathrm{YM}+\phi^3}(\{5\}|12[34]6)
\end{aligned} 
\end{equation}

\begin{figure}
    \centering
    \includegraphics[width=0.7\linewidth]{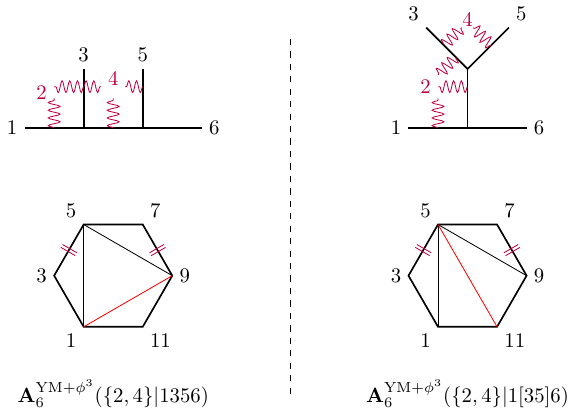}
    \caption{Double-gluon insertion at 6-point level, with inserted gluons being $\{2,4\}$.}
    \label{fig:gluon_Feyman3}
\end{figure}

The above procedure strongly suggests that $N_{n,r}$ depends only on the number of scalars, as it effectively treats all possible gluon insertions into a given scalar skeleton as a single collective object. For example, as shown in Figure~\ref{fig:gluon_Feyman3}, for $n=6$ with two gluons labeled by $2$ and $4$, the basis is given by:
\begin{equation}
\mathbf{A}_6^{\mathrm{YM}+\phi^3}(\{2,4\}|1356)\, ,  \mathbf{A}_6^{\mathrm{YM}+\phi^3}(\{2,4\}|1[35]6) \,.
\end{equation}

We also provide a proof of the statement~\eqref{eq: number_statement}, which can be summarized as follows. Given the pure scalar case, where $N_{n,n} = \mathcal{C}_{n-2}$ is trivial, we consider the insertion of a single gluon into the mixed amplitudes. We then show that the dimension of the basis for such amplitudes remains the same as that of the amplitudes prior to the gluon insertion, by proving that their all-order soft limits are uniquely determined by the pre-insertion amplitudes. This constitutes a generalized uniqueness of the mixed amplitudes, analogous to the uniqueness of pure YM amplitudes~\cite{Rodina:2016jyz}. Note that unlike pure YM theory which is uniquely determined by locality and gauge invariance, mixed amplitudes are determined by $\mathcal{C}_{r-2}$ independent parameters, modulo an overall scaling.

To conclude this section, we remark that it is natural to choose the mixed amplitudes with gluons inserted into a single scalar skeleton as the basis for the corresponding linear space. As we will see later, in certain cases these amplitudes admit a straightforward generalization, represented by a single multi-derivative operator acting on the YM amplitudes, analogous to the pure scalar case. Moreover, this basis naturally spans a planar expansion of the original Yang-Mills amplitudes.


\section{Gluon Insertions from Derivative}\label{sec: gluon insertion}

We have seen that the number of linearly independent building blocks for gluon and $\phi^3$ scalar mixed amplitudes with $r$ ($r\geq 2$) scalars is given by the Catalan number $\mathcal{C}_{r-2}$, which notably does not depend on the number of gluons. The linear space of mixed amplitudes with given gluons and scalars in different ordering is therefore spanned by the amplitudes of gluons inserted into single scalar diagram. This naturally leads to the question: can these amplitudes be obtained from pure Yang-Mills amplitudes by acting with suitable differential operators?

A naive approach would be to apply the chain rule to transform the operators in~\eqref{eq: ep operator}. However, this clearly results in linear combinations of multi-derivatives. It is therefore more desirable to study the behavior of single-term multi-derivatives, as they not only establish a direct connection between the pure gluon and mixed amplitudes, but also provide clearer structural control over the pure Yang-Mills amplitudes themselves. In this section, we present a simple answer: by following exactly the same rules as before (but with even fewer derivatives), one can obtain mixed amplitudes with gluons inserted into the scalar skeleton in infinitely many cases, while other types fail due to the violation of gauge invariance.


Starting with the pure YM amplitudes, it is clear that one needs to avoid derivatives involving $X_{2a,b}$ in order to preserve gluon $a$. In fact, the rule can be summarized as: 

\textit{Given gluons labeled by $a_1, a_2, \ldots, a_s$, the derivatives that extracted correponding mixed amplitudes are determined by the rules introduced in Section~\ref{sec: derivative to cubic}, applied to the subpolygon obtained by deleting vertices $2a_1-1, 2a_1, 2a_2-1, 2a_2, \ldots, 2a_s-1, 2a_s$ from the original configuration.} 

The single scalar diagram we choose to obtain the operators in this smaller problem is precisely the skeleton into which we insert the gluons. Importantly, as we will explain later, preserving gauge invariance sometimes requires a {\bf cyclic rotation} of the operators. Note that since there are exactly $\mathcal{C}_{r-2}$ such diagrams (derivatives) on the subpolygon, this gives the number of independent basis elements!
\begin{figure}
    \centering
    \includegraphics[width=0.55\linewidth]{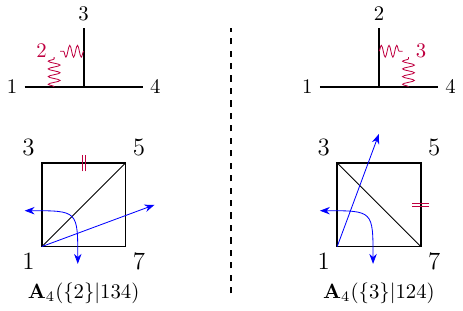}
    \caption{4-point gluon insertions shown on Feynman diagrams and their dual polygons. The blue-arrowed curves denote the corresponding derivatives acing on $\mathcal{A}_4^{\rm YM}$.}
    \label{fig:gluon1}
\end{figure}
Let $\mathbf{A}_n^{\mathrm{YM}+\phi^3}(\{a_1,\ldots,a_s\}| g)$ denote the $n-$point mixed amplitude with gluons ${a_1,\ldots,a_s}$ inserted into the scalar diagram $g$, where the 2-scalar cases are denoted by $\mathbf{A}_n^{\mathrm{YM}+\phi^3}(\{2,3,\ldots,n-1\}|1n)$ ~\footnote{For our purposes, when considering scalar amplitudes, we fix legs $1$ and $n$ to be scalars.}. We now provide examples to better illustrate this construction. For $n=4$, aside from the pure gluon and pure scalar cases, we have the mixed cases with $r=2$ and $r=3$. For the former, the amplitude with particles $1$ and $4$ being scalars is given by $\mathbf{A}_4^{\mathrm{YM}+\phi^3}(\{2,3\}|14)=\mathcal{D}_{(2,8)} \mathcal{A}_4^{\rm YM}$. For the latter, if particle $2$ is a gluon, we consider an effective 3-point problem labeled by vertices $1,2,5,6,7,8$  See figure~\ref{fig:gluon1}, and thus
\begin{equation}
    \mathbf{A}_4^{\mathrm{YM}+\phi^3}(\{2\}|134)= \mathcal{D}_{(2,8)(1,6)} \mathcal{A}_4^{\rm YM} \,,
\end{equation}
and similarly
\begin{equation}
    \mathbf{A}_4^{\mathrm{YM}+\phi^3}(\{3\}|124)= \mathcal{D}_{(2,8)(1,4)} \mathcal{A}_4^{\rm YM} \,.
\end{equation}
It is clear that for general multiplicity with $r=2,3$ we have:
\begin{equation}
    \mathbf{A}_n^{\mathrm{YM}+\phi^3}(\{2,3,\ldots,n-1\}|1n)=\mathcal{D}_{(2,2n)} \mathcal{A}_n^{\rm YM}\,,
\end{equation}
and 
\begin{equation}
    \mathbf{A}_n^{\mathrm{YM}+\phi^3}(\{2,3,\ldots,i-1,i+1,\ldots,n-1\}|1in)=\mathcal{D}_{(2,2n) (1,2i)} \mathcal{A}_n^{\rm YM}  \,.
\end{equation}
\begin{figure}
    \centering
    \includegraphics[width=0.95\linewidth]{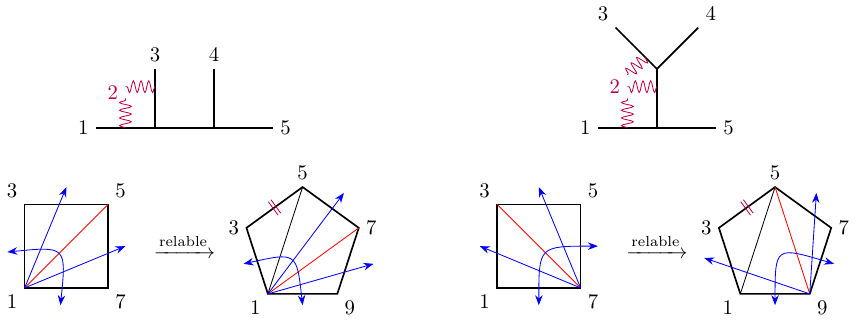}
    \caption{A 5-point mixed amplitude regarding ``2'' as the inserted gluon. The graphs on the left and right are taken from the contributions of 4-point $s$- and $t$-channels respectively.}
    \label{fig:gluon2}
\end{figure}
Starting from 5-point, we encounter more nontrivial cases with $r=4$. For instance, taking particle $2$ to be the gluon, the effective scalar problem is labeled by $1,2,5,\ldots,10$ See figure~\ref{fig:gluon2}. In this case, there are two contributing cubic diagrams, and the amplitudes are given by:
\begin{equation}
    \mathbf{A}_5^{\mathrm{YM}+\phi^3}(\{2\}|1345)= \mathcal{D}_{(2,10)(1,6)(1,8)} \mathcal{A}_5^{\rm YM} \,,
\end{equation}
where the operator $\mathcal{D}_{(2,10)(1,6)(1,8)}$ is a relabeling of $\mathcal{D}_{(2,8)(1,4)(1,6)}$. For diagram with propagator $1/X_{5,9}$ naively one expects the derivative to be $\mathcal{D}_{(2,10)(1,6)(5,8)}$ as the relabeling of $\mathcal{D}_{(2,8)(1,4)(3,6)}$. However, as we will explain shortly, we need to take its cyclic rotation and write:
\begin{equation} \label{eq:cyclic_example}
    \mathbf{A}_5^{\mathrm{YM}+\phi^3}(\{2\}|1[34]5)= \mathcal{D}_{(8,10)(2,9)(6,9)} \mathcal{A}_5^{\rm YM} \,.
\end{equation}

\begin{figure}
    \centering
    \includegraphics[width=0.55\linewidth]{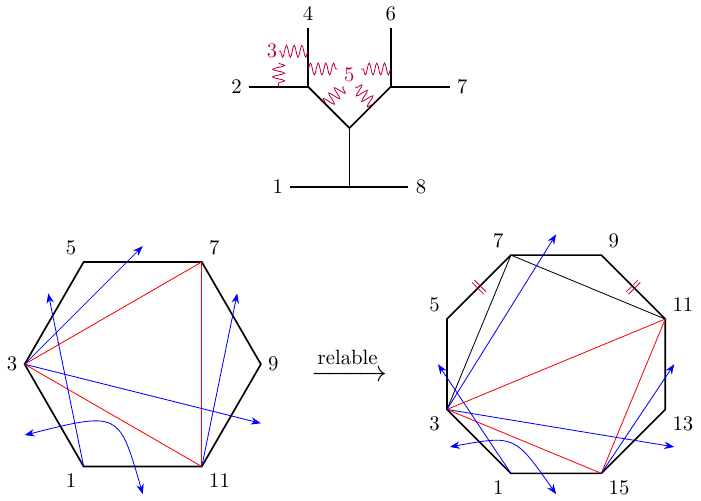}
    \caption{An 8-point example with gluons $\{3,5\}$ inserted into a 6-point scalar graph.}
    \label{fig:gluon3}
\end{figure}

Let us conclude by considering an 8-point example with 2 gluons and 6 scalars:
\begin{equation}
    \mathbf{A}_8^{\mathrm{YM}+\phi^3}(\{3,5\}|1[24][67]8)= \mathcal{D}_{(2,16)(1,4)(3,14)(15,12)(8,3)} \mathcal{A}_8^{\rm YM} \,.
\end{equation}
where the operator above is obtained by relabeling of $\mathcal{D}_{(2,12)(1,4)(3,10)(11,8)(6,3)}$ from $1,2,\ldots,12$ to $1,2,3,4,7,8,11,12,\ldots,15,16$ See figure~\ref{fig:gluon3}.
Next we provide strong consistency check by  proving certain operators obtained this way in general multiplicity is gauge invariant (and multi-linear) in remaining gluons. For simplicity we define:
\begin{equation}
    \mathcal{W}_{2i} \equiv \sum_{j\neq i-1,i,i+1} (X_{2i,j}-X_{2i-1,j}) \mathcal{D}_{(2i,j)} \,,
\end{equation}
and
\begin{equation}
    \tilde{\mathcal{W}}_{2i} \equiv \sum_{j\neq i-1,i,i+1} (X_{2i,j}-X_{2i+1,j}) \mathcal{D}_{(2i,j)} \,.
\end{equation}
Thus the gauge invariance and multi-linearity conditions~\eqref{eq:gauge1} and \eqref{eq:gauge2} read~\footnote{In the following we simply refer to gauge invariance and multi-linearity as gauge invariance.}:
\begin{equation}  \label{eq:gauge_operator}
    \mathcal{W}_{2i} \,\mathcal{A}_n^{\rm YM} = \tilde{\mathcal{W}}_{2i} \,\mathcal{A}_n^{\rm YM}=  \mathcal{A}_n^{\rm YM}\,.
\end{equation}
Given an operator $\mathcal{D}_\rho$ that does \textbf{not} involve the even labels $2 i_1, 2i_2,\ldots,2 i_s$, it preserves gauge invariance if:
\begin{equation}
   \mathcal{W}_{2i}  \mathcal{D}_\rho  \, \mathcal{A}_n^{\rm YM} - \mathcal{D}_\rho  \,\mathcal{A}_n^{\rm YM}=0\,,
\end{equation}
and 
\begin{equation}
     \tilde{\mathcal{W}}_{2i} \mathcal{D}_\rho  \, \mathcal{A}_n^{\rm YM} - \mathcal{D}_\rho \mathcal{A}_n^{\rm YM}=0\, ,
\end{equation}
for $i=i_1,i_2,\ldots,i_s$. Using~\eqref{eq:gauge_operator} the above conditions are equivalent to requiring the following commutators vanish:
\begin{equation}  \label{eq:gauge_commutator}[\mathcal{W}_{2i},\mathcal{D}_\rho]=[\tilde{\mathcal{W}}_{2i},\mathcal{D}_\rho]=0\,.
\end{equation}
It is easy to verify that for $\rho = (2a, 2b)$, equation~\eqref{eq:gauge_commutator} holds for any $i \neq a, b$. Naturally, no operator can extract an object that is gauge invariant at $(n-1)$ points. Now consider $\rho = (2a, 2b-1)$ and choose $i = b$, the commutator yields:
\begin{align}
    [\mathcal{W}_{2b},\mathcal{D}_{(2a,2b-1)}]&= (X_{2a,2b}-X_{2a,2b-1}) \mathcal{D}_{(2a,2b)} \mathcal{D}_{(2a,2b-1)}- \mathcal{D}_{(2a,2b)} (X_{2a,2b}-X_{2a,2b-1}) \mathcal{D}_{(2a,2b-1)} \nonumber \\
    &=-\mathcal{D}_{(2a,2b-1)} \, .
\end{align}
In the first line, we have used the fact that only the term with $j = 2a$ survives in the summation within $\mathcal{W}_{2b}$. And in the second line, we have used the linearity in the even label $2a$, as given in equation~\eqref{eq:labelMulti}. In fact, it is straightforward to find:
\begin{equation}
    [\mathcal{W}_{2i},\mathcal{D}_{(2a,2b-1)}]= -\delta_{i,b} \mathcal{D}_{(2a,2b-1)} \,,
\end{equation}
\begin{equation}[\mathcal{\tilde{W}}_{2i},\mathcal{D}_{(2a,2b+1)}]= -\delta_{i,b} \mathcal{D}_{(2a,2b+1)} \,.
\end{equation}
Therefore, for a derivative of the form $\mathcal{D}_{(2a,2b-1)}$, which involves one even label $2a$ and one odd label $2b-1$, the gauge invariance associated with particles $b$ and $b-1$ corresponding to even labels $2b$ and $2b-2$ is broken. This can be compensated by introducing $\mathcal{D}_{(2b,2b-2)}$ to effectively ``remove'' these gluons. This is precisely the mechanism used in constructing the 3-scalar amplitude with derivatives $\mathcal{D}_{(2,2n),(1,2a)}$, where $2$ and $2n$ are adjacent to $1$. And more generally, we have:
\begin{equation}
[\mathcal{W}_{2i},\mathcal{D}_{(2a',2b'),(2a_1,2b_1-1),\ldots,(2a_{m},2b_{m}-1)}]=[\mathcal{\tilde{W}}_{2i},\mathcal{D}_{(2a',2b'),(2a_1,2b_1-1),\ldots,(2a_{m},2b_{m}-1)}]=0\,.
\end{equation}
for all $i\neq a',b', a_1,a_2,\ldots,a_m$ if
\begin{equation}
(2b_1-1)\pm1,(2b_2-1)\pm1,\ldots,(2b_m-1)\pm1 \in \{2a',2b',2a_1,2a_2,\ldots,2a_m\} \,.
\end{equation}
In other words, every label adjacent to an odd index in the derivative must also appear as an even label within it. 

Now, returning to the example~\eqref{eq:cyclic_example} discussed earlier, we see that the derivative $\mathcal{D}_{(8,10)(2,9)(6,9)}$ satisfies the condition, since the labels $8$ and $10$, which are adjacent to $9$, appear in the derivative. In contrast, the expression before the cyclic rotation, namely $\mathcal{D}_{(2,10)(1,6)(5,8)}$, does not satisfy the condition: here, the presence of odd labels $1$ and $5$ requires that their adjacent even labels $2, 10, 4, 6$ all appear, however, label $4$ is missing.

\begin{figure}
    \centering
    \includegraphics[width=0.33\linewidth]{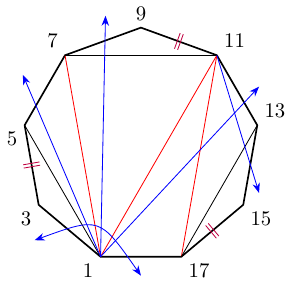}
    \caption{A 9-point mixed amplitude with gluons $\{2, 4, 6\}$ inserted into the scalar diagram $1235[78]9$ along with its corresponding derivative fails to preserve the gauge invariance of gluon $6$. }
    \label{fig:counter_example}
\end{figure}

It is easy to verify that our construction works for all $r=2,3,\ldots,n$ up to $n=8$. However, the first counterexamples appear at $n=9$. For instance, consider inserting gluons $2,5,8$ into the scalar diagram $134[67]9$, as shown in Figure~\ref{fig:counter_example}. The best one can do in this case is to construct the operator $\mathcal{D}_{(2,18)(1,6)(1,8)(1,12)(11,14)}$. However, gauge invariance for gluon $6$ is broken due to the presence of the derivative $\partial/\partial X_{11,14}$. In fact, we have:
\begin{equation}
    [\mathcal{W}_{2i},\mathcal{D}_{(2,18)(1,6)(1,8)(1,12)(11,14)}]=-\mathcal{D}_{(2,18)(1,6)(1,8)(1,12)(12,14)} \,.
\end{equation}

It would be interesting to study general operators, or combinations thereof, that preserve gauge invariance in this context; we leave this investigation for future work.

\section{The Planar Universal Expansion}\label{sec: universal expansion}
The lesson from the previous sections is that the optimal asymptotic behavior of the number of terms in an expansion for gluon amplitudes should not go beyond Catalan number. A natural application of this observation is to rewrite the universal expansion~\eqref{eq:expansion} in terms of the basis we have discussed:
\begin{equation} \label{eq: planar universal expansion}
\begin{aligned}
    {\cal A}^\text{YM}_n
    = \sum_{m,1 g_{\{\alpha\}}n } (-1)^m \, W(1\; g_{\{\alpha\}} \; n)  \, {\bf A}_n^{\mathrm{YM}+\phi^3}(\{\bar{\alpha}\}|  1 g_{\{\alpha\}} n )\,,
\end{aligned}
\end{equation}
where the sum runs over all unordered subsets ${\alpha}$ of ${2,3,\ldots,n}$ with $|\alpha|=m$. For each subset, we further sum over $\mathcal{C}_m$ scalar cubic graphs $1 g_{\{\alpha\}}  n$, labeled by commutators that correspond to subtrees along the $(1,n)$ line. In this rewritten form, the original prefactor acquires precisely the same commutator structure (c.f.~\cite{Bern:2011ia}) and the new prefactor is denoted by $W(1\; g_{\{\alpha\}}\;n)$. For example:
\begin{equation*}
\begin{aligned}
    W(123456)=\epsilon_1\cdot f_2\cdot f_3\cdot f_4\cdot f_5\cdot\epsilon_6, \quad &W(1[23]456)=\epsilon_1\cdot f_{[2}\cdot f_{3]}\cdot f_4\cdot f_5\cdot\epsilon_6,  \\
     W(1[[23]4]56)=\epsilon_1\cdot f_{[[2}\cdot f_{3]}\cdot f_{4]}\cdot f_5\cdot\epsilon_6, \quad &W(1[[23][45]]6)=\epsilon_1\cdot f_{[[2}\cdot f_{3]}\cdot f_{[4}\cdot f_{5]]}\cdot\epsilon_6 \,.
\end{aligned}
\end{equation*}
One can of course express the prefactor in terms of the scalar-scaffolded variables. Note that~\cite{Cao:2025lzv}:
\begin{equation}
f_{i}^{\mu \nu}=k_{i}^{[\mu}\epsilon_{i}^{\nu]}=-(p_{2i-1}+p_{2i})^{[\mu}p_{2i-1}^{\nu]}=p_{2i-1}^{[\mu} p_{2i}^{\nu]} \,,
\end{equation}
where we have used the gauge choice $\lambda=0$ in~\eqref{eq:polarization to 2n-gon}~\footnote{In our convention, we take $\epsilon_i = -p_{2i-1}$, whereas Ref.~\cite{Cao:2025lzv} adopts $\epsilon_i = p_{2i-1}$.}. We can then express the original prefactor as the following commutator:
\begin{equation}
    \epsilon_1\cdot f_{\alpha_1}\cdot f_{\alpha_2}\cdots f_{\alpha_m}\cdot\epsilon_n= s_{1,[2\alpha_1-1} s_{2\alpha_1+1],[2\alpha_2-1} s_{2\alpha_2],[2\alpha_3-1} \ldots s_{2\alpha_m],2n-1} ,
\end{equation}
where $s_{i,j}=(p_i+p_j)^2 =2\,p_i \cdot p_j$. We have omitted a factor of $(1/2)^{2m+1}$ as well as that could appear in the mixed amplitude, which can be absorbed into the definition of the coupling constant. The prefactor $W(1, g_{{\alpha}}, n)$ is then constructed as a nested commutator built upon this structure. Let us present the explicit expansion for $n=4,5$. 

For $n=4$, the original amplitudes are reorganized as:
\begin{align}
    \mathcal{A}_4^{\mathrm{YM}+\phi^3}(\{2,3\}|1,4)&=\mathbf{A}_4^{\mathrm{YM}+\phi^3}(\{2,3\}|14),\\
    \mathcal{A}_4^{\mathrm{YM}+\phi^3}(\{3\}|1,2,4)&=\mathbf{A}_4^{\mathrm{YM}+\phi^3}(\{3\}|124),\\
    \mathcal{A}_4^{\mathrm{YM}+\phi^3}(\{2\}|1,3,4)&=\mathbf{A}_4^{\mathrm{YM}+\phi^3}(\{2\}|134),\\
    \mathcal{A}_4^{\mathrm{YM}+\phi^3}(\varnothing|1,2,3,4)&=\mathbf{A}_4^{\mathrm{YM}+\phi^3}(\varnothing |1234)+\mathbf{A}_4^{\mathrm{YM}+\phi^3}(\varnothing |1[23]4),\\
    \mathcal{A}_4^{\mathrm{YM}+\phi^3}(\varnothing|1,3,2,4)&=-\mathbf{A}_4^{\mathrm{YM}+\phi^3}(\varnothing |1[23]4).
\end{align}
Plugging the above results into~\eqref{eq:4pt universal expansion} we find:
\begin{equation}
     \begin{aligned}
        \mathcal{A} _{4}^{\mathrm{YM}}&=W\left( 14 \right) \mathbf{A}_4^{\mathrm{YM}+\phi^3}(\{2,3\}|14)\\&-W\left( 124 \right) \mathbf{A}_4^{\mathrm{YM}+\phi^3}(\{3\}|124)-W\left( 134 \right) \mathbf{A}_4^{\mathrm{YM}+\phi^3}(\{2\}|134)\\ &    +W\left( 1234 \right) \mathbf{A}_4^{\mathrm{YM}+\phi^3}(\varnothing |1234)+W\left( 1[23]4 \right) \mathbf{A}_4^{\mathrm{YM}+\phi^3}(\varnothing |1[23]4).
    \end{aligned}
\end{equation}
Similarly, for $n=5$ we have (see Appendix~\ref{app: expansion} for details):
\begin{equation} \label{eq: 5pt planar universal expansion}
     \begin{aligned}
        \mathcal{A} _{5}^{\mathrm{YM}}&=W\left( 15 \right) \mathbf{A}_5^{\mathrm{YM}+\phi^3}\left( \left\{ 2,3,4 \right\} |15 \right) 
-W\left( 125 \right) \mathbf{A}_5^{\mathrm{YM}+\phi^3}\left( \left\{ 3,4 \right\} |125 \right) \\&-W\left( 135 \right) \mathbf{A}_5^{\mathrm{YM}+\phi^3}\left( \left\{ 2,4 \right\} |135 \right) -W\left( 145 \right) \mathbf{A}_5^{\mathrm{YM}+\phi^3}\left( \left\{ 2,3 \right\} |145 \right) 
\\
& +W\left( 1235 \right) \mathbf{A}_5^{\mathrm{YM}+\phi^3}\left( \left\{ 4 \right\} |1235 \right) +W\left( 1[23] 5 \right) \mathbf{A}_5^{\mathrm{YM}+\phi^3}\left( \left\{ 4 \right\} |1[23] 5 \right) 
\\
&+W\left( 1245 \right) \mathbf{A}_5^{\mathrm{YM}+\phi^3}\left( \left\{ 3 \right\} |1245 \right) +W\left( 1[24] 5 \right) \mathbf{A}_5^{\mathrm{YM}+\phi^3}\left( \left\{ 3 \right\} |1[24] 5 \right) 
\\
& +W\left( 1345 \right) \mathbf{A}_5^{\mathrm{YM}+\phi^3}\left( \left\{ 2 \right\} |1345 \right) +W\left( 1[34] 5 \right) \mathbf{A}_5^{\mathrm{YM}+\phi^3}\left( \left\{ 2 \right\} |1[34] 5 \right) 
\\
& -W\left( 12345 \right) \mathbf{A}_5^{\mathrm{YM}+\phi^3}\left( \varnothing |12345 \right) -W\left( 1[[23] 4] 5 \right) \mathbf{A}_5^{\mathrm{YM}+\phi^3}\left( \varnothing |1[[23] 4] 5 \right) 
\\
& -W\left( 12[34] 5 \right) \mathbf{A}_5^{\mathrm{YM}+\phi^3}\left( \varnothing |12[34] 5 \right) -W\left( 1[23] 45 \right) \mathbf{A}_5^{\mathrm{YM}+\phi^3}\left( \varnothing |1[23] 45 \right) \\&-W\left( 1[2[34]] 5 \right) \mathbf{A}_5^{\mathrm{YM}+\phi^3}\left( \varnothing |1[2[34]] 5 \right) .
    \end{aligned}
\end{equation}

Note that the number of terms in this planar universal expansion is given by:
\begin{equation}
\sum_{m=0}^{n-2} \mathcal{C}_m \binom{n-2}{m} ,
\end{equation}
which takes the values $2, 5, 15, 51, 188, 731$ for $n=3,4,5,6,7,8$, and asymptotically grows as $5^{n+1/2}/(8\sqrt{\pi} \,n^{3/2})$ for large $n$ (see (\href{https://oeis.org/A007317}{OEIS A007317})). 

It would be interesting to further investigate the expansion of the mixed amplitudes within the planar universal expansion. If one follows the reference ordering dependent approach of~\cite{Teng:2017tbo,Fu:2017uzt}, the final result, of course, reproduces their expressions, yielding the (commutators of) BCJ numerators multiplied by pure scalar diagrams. However, we remark that our formulation involves fewer terms at each intermediate step. For example, for the $(n-2)$-gluon amplitudes, one finds:
\begin{equation}
\begin{aligned}
    &{\cal A}_n^{\mathrm{YM}+\phi^3}(\{2,3,\ldots,n-1\}|  1 n )={\bf A}_n^{\mathrm{YM}+\phi^3}(\{2,3,\ldots,n-1\}|  1 n )\\
    =& \sum_{s,\beta}\ \tilde{W}(\beta ;2  )\  {\cal A}_n^{\mathrm{YM}+\phi^3}(\{\bar{\beta}\}|  1 \beta^{-1}  2 n )\,,
\end{aligned}
\end{equation}
where further choose leg $2$ to be special and sum over the ordered subsets $\beta$ with $|\beta|=s$ and $\beta^{-1}$ to be its reverse ordered set. The prefactor $\tilde{W}(\beta; 2  )$ is defined by
\begin{equation}
    \tilde{W}(\beta; 2 )= \epsilon_2 \cdot f_{\beta_1}\cdot f_{\beta_2} \cdots  f_{\beta_s} \cdot k_1\,,
\end{equation}
or equivalently
\begin{equation}
\begin{aligned}
    \tilde{W}(\beta; 2 )&= s_{1,[2\beta_1-1} s_{2\beta_1+1],[2\beta_2-1} s_{2\beta_2],[2\beta_3-1} \ldots s_{2\beta_s],3}\\
    &+s_{2,[2\beta_1-1} s_{2\beta_1+1],[2\beta_2-1} s_{2\beta_2],[2\beta_3-1} \ldots s_{2\beta_s],3} 
\end{aligned}
\end{equation}
in the scalar scaffolded kinematics. Rewriting the above expression in our basis gives:
\begin{equation}
  {\bf A}_n^{\mathrm{YM}+\phi^3}(\{2,3,\ldots,n-1\}|  1 n )
= \sum_{s,1g_{(2\beta)}n}(-1)^{s}\ \tilde{W}(g_{(\beta)};2  )\  {\bf A}_n^{\mathrm{YM}+\phi^3}(\{\bar{\beta}\}|  1g_{(2\beta )}n )\,,
\end{equation}
where $1g_{(2\beta)}n$ denotes the graphs corresponding to the full commutator that includes $2$ and $\beta$, while $g_{(\beta)}$ in $g_{(\beta)};2$ represents the associated commutator without $2$. For instance, the 5-point 3 gluons amplitude is given by:
\begin{equation}
\begin{aligned}
{\bf A}_n^{\mathrm{YM}+\phi^3}(\{2,3,4\}|  1 5 )&=\tilde{W}(\varnothing;2) {\bf A}_n^{\mathrm{YM}+\phi^3}(\{34\}|  125 )\\
&-\tilde{W}(3;2){\bf A}_n^{\mathrm{YM}+\phi^3}(\{4\}|  1[23]5 )-\tilde{W}(4;2){\bf A}_n^{\mathrm{YM}+\phi^3}(\{3\}|  1[24]5 )
\\
&+ \tilde{W}(34;2){\bf A}_n^{\mathrm{YM}+\phi^3}(\varnothing|  1[[23]4]5 )+\tilde{W}([34];2){\bf A}_n^{\mathrm{YM}+\phi^3}(\varnothing|  1[2[34]]5 )\,.
\end{aligned}
\end{equation}
We expect to develope a different approach that eliminates reference-ordering dependence while manifestly preserving the planar structure for both pure YM and mixed amplitudes. We leave this for future work.

\section{Conclusions and Outlook}\label{sec: conclusion}
In this work, we have revisited and explored some structural aspects of Yang-Mills amplitudes written in scaffolding variables. By analyzing their factorizations, we developed a systematic prescription for constructing differential operators that map the gluon amplitudes into single cubic scalar diagrams. We also considered a naive generalization of this construction to obtain the mixed amplitudes of gluons and scalars, which succeeds/fails in infinite cases of given types. We showed that the mixed amplitude with $r$ scalars and any number of gluons admits a natural basis consisting of $\mathcal{C}_{r-2}$ elements. We then applied this result to construct a planar version of the universal expansion, in which each term consists of a gauge-invariant prefactor expressed as a nested commutator, accompanied by the mixed amplitudes defined in our basis.

There remains several directions for future investigation. First, it would be interesting to further study the structure of the differential operators that preserve gauge invariance in the scaffolding variables and present a complete set of differential operators that generate all mixed amplitudes in this planar basis. It would also be worthwhile to study the hidden zeros, factorizations near zeros, and the splittings~\cite{Arkani-Hamed:2023swr, Cao:2024gln, Bartsch:2024amu, Li:2024qfp,Arkani-Hamed:2024fyd} of such mixed amplitudes, as well as their relation to these operators. In addition, developing new approach that fully manifests the planar structure for mixed amplitudes would represent a natural next step toward a more universal understanding of YM amplitudes (see~\cite{Backus:2025orx} for a recent paper on new recursion of scalar-scaffolded pure YM). 

Moreover, our results hint at broader applicability beyond YM theory. Since all colored amplitudes, such as those in the Nonlinear Sigma Model, colored Yukawa theory~\cite{De:2024wsy}, and even certain subsets of the mixed amplitudes involving gravitons, gluons, and scalars, exhibit similar structures~\cite{Arkani-Hamed:2024vna}, understanding whether the differential-operator methods developed here can be extended to these theories would shed light on the universality of the scaffolded approach.

Another natural question is whether the same ideas apply to higher-dimensional deformations of Yang-Mills theory, and more generally to full string amplitudes rather than only their field-theory limit. The differential operators studied here are close relatives of the scalar-scaffolded operators in~\cite{Backus:2025njt} and of the transmutation operators in~\cite{Cheung:2017ems}. The operators in~\cite{Backus:2025njt} lead to shifted stringy Tr$\phi^3$ objects whose field-theory limits give the ordinary Tr$\phi^3$ amplitudes, while the operators in~\cite{Cheung:2017ems} can be applied directly to string amplitudes, producing stringy ultraviolet completions of the corresponding mixed amplitudes. It is therefore natural to expect that the operators considered in the present paper can also be generalized to higher-derivative corrections and to full open-string amplitudes. However, the Catalan counting established in this paper relies on a uniqueness argument whose crucial input is the mass dimension of the minimal YM/YMS amplitudes. For higher-dimensional operators, and especially at finite $\alpha'$, the corresponding amplitudes involve additional numerator factors or stringy decorations with higher mass dimension. Moreover, already in the pure-scalar sector, the finite-$\alpha'$ objects are the $Z$-integrals~\cite{Carrasco:2016ldy,Mafra:2016mcc}, which do not admit the same Catalan basis as their field-theory limit. It would therefore be interesting to investigate, order by order in $\alpha'$, whether the structures found here nevertheless imply a basis smaller than the usual $(r{-}2)!$-dimensional one.

Finally, it would be very interesting to extend our analysis beyond tree level~\cite{Zhou:2022djx,Chen:2023bji} or into theories with richer matter content (c.f.~\cite{Balli:2024wje, Geyer:2024oeu, Cao:2024olg,Cao:2025ygu,Monteiro:2025qai, Du:2025yxz} for recent works on one-loop field theory and string amplitudes). An immediate next step is to investigate whether an analogue of our  planar expansion exists at one loop~\cite{Cao:2024olg}.


\acknowledgments
It is our pleasure to thank Song He for stimulating discussions, encouragement and collaborations on the early stage of this project. We also thank Qu Cao and Fan Zhu for discussions. The work of JD is supported by the DFG grant 508889767,
Forschungsgruppe ``Modern foundations of scattering amplitudes''.

\appendix

\section{Details of Minimal Basis for Mixed Amplitudes} \label{app: rank}
\subsection{6-point 2-gluon examples} \label{app: rank example}
We provide an explicit example to demonstrate that $N_{6,5}=5$. Without loss of generality, we choose the only gluon to be $5$ and define:
\begin{equation}
\begin{aligned}
    &\mathcal{A}_6^{\mathrm{YM}+\phi^3}(\{5\}|\beta)\\
    =& n(\beta) \cdot \left(\frac{X_{7,10}}{X_{3,11} X_{5,11} X_{7,11}},\frac{X_{7,10}}{X_{1,5} X_{5,11} X_{7,11}},\frac{X_{7,10}}{X_{3,7} X_{3,11} X_{7,11}},\frac{X_{7,10}}{X_{1,7} X_{3,7} X_{7,11}},\frac{X_{7,10}}{X_{1,5} X_{1,7} X_{7,11}}, \right.\\
    &\left.
    \frac{X_{5,10}}{X_{3,11} X_{5,9} X_{5,11}},  \frac{X_{5,10}}{X_{1,5} X_{5,9} X_{5,11}},\frac{X_{3,10}}{X_{3,9} X_{3,11} X_{5,9}},\frac{X_{1,10}}{X_{1,9} X_{3,9} X_{5,9}},\frac{X_{1,10}}{X_{1,5} X_{1,9} X_{5,9}},\frac{X_{3,10}}{X_{3,7} X_{3,9} X_{3,11}},\right.\\
    &\left.
    \frac{X_{1,10}}{X_{1,9} X_{3,7} X_{3,9}},\frac{X_{1,10}}{X_{1,7} X_{1,9} X_{3,7}},\frac{X_{1,10}}{X_{1,5} X_{1,7} X_{1,9}},\frac{1}{X_{3,11} X_{5,11}},\frac{1}{X_{1,5} X_{5,11}},\frac{1}{X_{3,11} X_{5,9}},
    \right.\\ &\left.
    \frac{1}{X_{3,9} X_{5,9}},\frac{1}{X_{1,5} X_{5,9}},\frac{1}{X_{3,7} X_{3,11}},\frac{1}{X_{3,7} X_{3,9}},\frac{1}{X_{1,7} X_{3,7}},\frac{1}{X_{1,5} X_{1,7}}\right)\,.
\end{aligned} 
\end{equation}
Note that in this case there are $12$ cyclic and reflection inequivalent orderings, namely:
\begin{equation}
\begin{aligned}
 \{ & (1,2,3,4,6),(1,2,3,6,4),(1,2,4,3,6),(1,2,4,6,3),(1,2,6,3,4),(1,2,6,4,3),\\
 &(1,3,2,4,6),(1,3,2,6,4),(1,3,4,2,6),(1,3,6,2,4),(1,4,2,3,6),(1,4,3,2,6)\}\,.
\end{aligned}
\end{equation}
The corresponding $12 \times 23$ matrix is then given by:
\begin{equation*}
    \left(
\begin{array}{ccccccccccccccccccccccc}
 -1 & -1 & -1 & -1 & -1 & -1 & -1 & -1 & -1 & -1 & -1 & -1 & -1 & -1 & 1 & 1 & 1 & 1 & 1 & 1 & 1 & 1 & 1 \\
 0 & 0 & 0 & 1 & 1 & 0 & 0 & 0 & 0 & 0 & 0 & 0 & 1 & 1 & 0 & 0 & 0 & 0 & 0 & 0 & 0 & -1 & -1 \\
 1 & 1 & 0 & 0 & 0 & 1 & 1 & 1 & 1 & 1 & 0 & 0 & 0 & 0 & -1 & -1 & -1 & -1 & -1 & 0 & 0 & 0 & 0 \\
 0 & 0 & 0 & 0 & 1 & 0 & 0 & 0 & 0 & 0 & 0 & 0 & 0 & 1 & 0 & 0 & 0 & 0 & 0 & 0 & 0 & 0 & -1 \\
 0 & 1 & 0 & 0 & 0 & 0 & 1 & 0 & 0 & 1 & 0 & 0 & 0 & 0 & 0 & -1 & 0 & 0 & -1 & 0 & 0 & 0 & 0 \\
 0 & -1 & 0 & 0 & -1 & 0 & -1 & 0 & 0 & -1 & 0 & 0 & 0 & -1 & 0 & 1 & 0 & 0 & 1 & 0 & 0 & 0 & 1 \\
 0 & 0 & 1 & 1 & 0 & 0 & 0 & 0 & 0 & 0 & 1 & 1 & 1 & 0 & 0 & 0 & 0 & 0 & 0 & -1 & -1 & -1 & 0 \\
 0 & 0 & 0 & -1 & 0 & 0 & 0 & 0 & 0 & 0 & 0 & 0 & -1 & 0 & 0 & 0 & 0 & 0 & 0 & 0 & 0 & 1 & 0 \\
 1 & 0 & 0 & 0 & 0 & 1 & 0 & 1 & 1 & 0 & 0 & 0 & 0 & 0 & -1 & 0 & -1 & -1 & 0 & 0 & 0 & 0 & 0 \\
 0 & 0 & 0 & 0 & 0 & 0 & 0 & 0 & 0 & 0 & 0 & 0 & 0 & 0 & 0 & 0 & 0 & 0 & 0 & 0 & 0 & 0 & 0 \\
 0 & 0 & 1 & 0 & 0 & 0 & 0 & 0 & 0 & 0 & 1 & 1 & 0 & 0 & 0 & 0 & 0 & 0 & 0 & -1 & -1 & 0 & 0 \\
 -1 & 0 & -1 & 0 & 0 & -1 & 0 & -1 & -1 & 0 & -1 & -1 & 0 & 0 & 1 & 0 & 1 & 1 & 0 & 1 & 1 & 0 & 0 \\
\end{array}
\right) 
\end{equation*}
and one can readily verify that this matrix has rank $5$.

\subsection{Proof} \label{app: proof}
In this appendix, we prove the number of linearly independent mixed amplitudes with $r$ scalars and $n-r$ gluons is $\mathcal{C}_{r-2}$
We start with an ansatz keeping identical propagators with the mixed amplitudes (i.e. obeying locality), denoted as
\begin{equation}
    M_n(p^l)=\sum_{g}\frac{\mathcal{M}_g(p^l)}{\prod_{i}P_{g,i}},
\end{equation}
in which the sum is over different Feynman diagrams $g$, and $\mathcal M_g(p^l)$ is a polynomial yet to be determined with $l$-power in momentum. We could further determine the power of momentum is counting mass dimensions. Denote the operator ``$[\cdot]$'' for counting the power of momenta in the numerator (abbreviated for ``PMN'' in the following context) for a given quantity. We then implement these operators on equation \eqref{eq:expansion}, which in turn gives us
\begin{equation}
    [\mathcal{A}^{\mathrm{YM}}_n]=[\epsilon_1\cdot f_{\alpha_1}\cdots f_{\alpha_{r-2}}\cdot\epsilon_n]+[\mathcal{A}_n^{\mathrm{YM}+\phi^3}(\{\bar\alpha\}|1,\alpha,n)].
\end{equation}
Since the PMN of $\epsilon_1\cdot f_{\alpha_1}\cdots f_{\alpha_{r-2}}\cdot\epsilon_n$ and $\mathcal{A}_n^{\rm YM}$ are respectively
\begin{equation}
    [\mathcal{A}_n^{\rm YM}]=n-2,\qquad[\epsilon_1\cdot f_{\alpha_1}\cdots f_{\alpha_{r-2}}\cdot\epsilon_n]=r-2,
\end{equation}
then the PMN for the mixed amplitude $\mathcal{A}_n^{\mathrm{YM}+\phi^3}(\{\bar\alpha\}|1,\alpha,n)$ is
\begin{equation}
    [\mathcal{A}_n^{\mathrm{YM}+\phi^3}(\{\bar\alpha\}|1,\alpha,n)]=n-r.
\end{equation}

Based on these analysis, the PMN of our anstaz is confirmed to be $n-r$, thus it becomes
\begin{equation}\label{eq:anstaz}
    M_n(p^{n-r})=\sum_g\frac{\mathcal{M}_g(p^{n-r})}{\prod_iP_{g,i}}.
\end{equation}
The idea is to show that inserting gluons will not affect the number of minimal basis. Or equivalent saying, if all terms in a $r$-scalar $\phi^3$ amplitude has been given, then no matter inserting how many external gluons, the mixed amplitude $\mathcal{A}_{n}^{\mathrm{YM}+\phi^3}(\{\bar\alpha\}|1,\alpha,n)$ can always be fixed by the $r$-scalar amplitude $\mathcal{A}_r^{\phi^3}(1,\alpha,n)$.
We will use mathematical induction in our proof, in which we admit that at $n$-point level with $r$ scalars insertion whose $n-r$ gluons satisfy gauge invariance, the mixed amplitude $\mathcal{A}_n^{\mathrm{YM}+\phi^3}(\{\bar\alpha\}|1,\alpha,n)$ can be uniquely determined by $r$-scalar amplitude $\mathcal{A}_r^{\phi^3}$ (or mixed amplitude $\mathcal{A}_3^{\mathrm{YM}+\phi^3}(2|1,3)$ in case of $r=2$), and the scalar amplitudes themselves can produce $\mathcal{C}_{r-2}$ minimal basis.

Now we insert a gluon external leg, with its momentum $p_{n+1}$. Denote
\begin{equation}
    p_{n+1}^\mu=zq^\mu.
\end{equation}
There are two possible ways of insertion that is consistent with Feynman rules:
\begin{figure}
    \centering
    \begin{minipage}[t]{0.35\textwidth}    \includegraphics[scale=1.2]{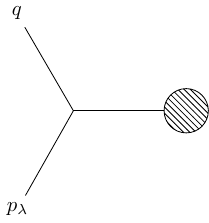}
    \caption{Gluon inserted on an external leg.}
    \label{fig:single pole insertion}
    \end{minipage}
    \hspace{0.1\textwidth}
    \begin{minipage}[t]{0.35\textwidth}
    \includegraphics[scale=1.2]{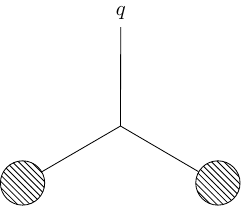}
    \caption{Gluon inserted on a propagator.}
    \label{fig:non-single pole insertion} 
    \end{minipage}
\end{figure}
\begin{enumerate}
    \item If the gluon (denoted as ``$q$'') is injected on the $\lambda$-th external leg (see fig.\ref{fig:single pole insertion}), this corresponds to the singular pole with the propagator $s_{\lambda,n+1}=2zp_\lambda\cdot q$, while on other propagators, $z$ only contributes to spurious poles such that no singularities occurs.
    \item If on the other hand, the gluon is inserted on a propagator between $\lambda$-th and $(\lambda+1)$-th external legs (see fig.\ref{fig:non-single pole insertion}), such diagram can be factored out as
    \begin{equation}
        \frac{N(z)}{P_L(z)(p_1+\cdots+p_\lambda)^2(zq+p_1+\cdots+p_\lambda)^2P_R(z)},
    \end{equation}
    where $P_L(z)$ is the product of propagators on the left of $\lambda$-th external leg and $P_R(z)$ denotes the product of propagators on the right of the $(\lambda+1)$-th external leg.
\end{enumerate}
Based on these discussions, we can now perform a soft-momentum expansion on our $(n+1)$-point ansatz:
\begin{equation}
    M_{n+1}=z^{\ell_{\min}} M_{n+1}^{(\ell_{\min})}+z^{\ell_{\min}+1} M_{n+1}^{(\ell_{\min}+1)}\cdots+z^{\ell_{\max}} M_{n+1}^{(\ell_{\max})}.
\end{equation}
Note that for the lowest degree $\ell_{\min}$, it can be given by one diagram has singular pole with numerator independent of $z$. This tells us that $\ell_{\min}=-1$. For the highest degree $\ell_{\max}$, since we have shown that the PMN of $M_{n+1}$ previously is $n+1-r$, then as an extreme, we expect all momenta contribute to it is just $q$ when the propagators are non-singular. This tells us $\ell_{\max}=n+1-r$. Thus the soft expansion can be written as
\begin{equation}
    M_{n+1}=\frac{M_{n+1}^{(-1)}}{z}+M_{n+1}^{(0)}+zM_{n+1}^{(1)}\cdots+z^{n+1-r} M_{n+1}^{(n+1-r)}.
\end{equation}
This indicates that we only need to prove our claim in each expansion order of $M_{n+1}$. We illustrate our idea in several steps:\footnote{Without loss of generality, we assume the color ordering of the $(n+1)$-point mixed amplitude to be $$\mathcal{A}_{n+1}^{\mathrm{YM}+\phi^3}(\underbrace{\{r,r+1\dots,n-1,n+1\}}_{n+1-r\text{ gluons}}|\underbrace{1,2,\dots,r-1,n}_{r\text{ scalars}})=:\mathcal{A}_{n+1}^{\mathrm{YM}+\phi^3}(r).$$}

At the leading order, we set $(n+1)$-th gluon injecting after the $n$-th particle. By this operation, we obtain a new ordering, say $\Tr(1,2,\dots,n+1)$. Note that the ordering obeys the $Z_{n+1}$ symmetry, which indicates that
\begin{equation}
    \Tr(1,2,\dots,n+1)=\Tr(n+1,1,2\dots,n).
\end{equation}
This shows us that inserting a gluon after the $n$-th particle is equivalent to insert it before the first particle, revealing that $M_{n+1}^{(-1)}$ has two singular poles
\begin{equation}
    M_{n+1}^{(-1)}(p^{n+1-r})=\frac{\epsilon_{n+1}^\mu\Gamma_\mu(p^{n+1-r})}{q\cdot p_1}+\frac{\epsilon_{n+1}^\mu\Delta_\mu(p^{n+1-r})}{q\cdot p_{n}}.
\end{equation}
Gauge invariance at $(n+1)$-th gluon requires that both $\Gamma_\mu(p^{n+1-r})$ and $\Delta_\mu(p^{n+1-r})$ can be factored out as 
\begin{equation}
    \Gamma_\mu(p^{n+1-r})=(p_1)_\mu\Gamma(p^{n-r}),\qquad\Delta_\mu(p^{n+1-r})=-(p_n)_{\mu}\Gamma(p^{n-r}),
\end{equation}
where $\Gamma(p^{n-r})$ is an arbitrary anstaz in terms of \eqref{eq:anstaz}. This leads to
\begin{equation}
    M_{n+1}^{(-1)}(p^{n+1-r})=\left(\frac{\epsilon_{n+1}\cdot p_1}{q\cdot p_1}-\frac{\epsilon_{n+1}\cdot p_n}{q\cdot p_n}\right)\Gamma(p^{n-r}).
\end{equation}
Note that since $\Gamma(p^{n-r})$ is gauge invariant in $n-r$ gluons, by our induction assumption, such anstaz can be fixed with $\mathcal{A}_{n}^{\mathrm{YM}+\phi^3}(r)$. In this way, we find that the leading order $M_{n+1}^{(-1)}$ is given by the leading order of $\mathcal{A}_{n+1}^{\mathrm{YM}+\phi^3}(r)$, {\it i.e.}
\begin{equation}
    M_{n+1}^{(-1)}(p^{n+1-r})=\left[\mathcal{A}_{n+1}^{\mathrm{YM}+\phi^3}(r)\right]^{(-1)}.
\end{equation}

At the sub-leading order, there are two types of denominator now: either linear in $q$ or being a polynomial of $q$. This gives us two types of terms in the following:
\begin{equation}\label{eq:subleading order}
    M_{n+1}^{(0)}(p^{n+1-r})=\sum_{i=1,n}\frac{\epsilon_{n+1}^\mu q^\nu\Gamma_{i;\mu\nu}(p^{n-r})}{q\cdot p_i}+\sum_{i=2}^{n-1}\frac{\epsilon_{n+1}^\mu\Delta_{i;\mu}(p^{n-r+1})}{P_i},
\end{equation}
where $P_i:=(p_1+\cdots+p_i)^2$. Note that since $\Delta_{i;\mu}(p^{n+1-r})$ is lack of the polarization vector, it can not achieve gauge invariance for $n-r+1$ gluons. However at the same time, it is required that $M_{n+1}^{(0)}(p^{n+1-r})$ should be gauge invariant in these gluons, hence the only choice left for $\Delta_{i;\mu}(p^{n+1-r})$ is to vanish. For terms associated with the singular poles, one can let 
\begin{equation}
    \Gamma_{n;\mu\nu}(p^{n-r})=-\frac{q\cdot p_{n}}{q\cdot p_1}\Gamma_{1;\mu\nu}(p^{n-r})
\end{equation}
to anti-symmetrize the prefactor in \eqref{eq:subleading order}:
\begin{equation}
    M_{n+1}^{(0)}(p^{n+1-r})=\epsilon_{n+1}^{[\mu}q^{\nu]}\Gamma_{1;\mu\nu}(p^{n-r}).
\end{equation}
Thus alike the case in the leading order term, $\Gamma_{1;\mu\nu}(p^{n-r})$ has maximally $n-r$ $\epsilon\cdot p$ terms and gauged in $n-r$ gluons, which can be again fixed by $\mathcal{A}_n^{\mathrm{YM}+\phi^3}(r)$. For this reason, we conclude that 
\begin{equation}
    M^{(0)}_{n+1}(p^{n+1-r})=\left[\mathcal{A}_{n+1}^{\mathrm{YM}+\phi^3}(r)\right]^{(0)}.
\end{equation}

At the sub-sub-leading order, analogously we have
\begin{equation}
    M_{n+1}^{(1)}(p^{n+1-r})=\epsilon_{n+1}^{[\mu}q^{\nu]}\sum_{i=1,n}\frac{q^\rho\Gamma_{i;\mu\nu\rho}(p^{n-1-r})}{q\cdot p_i}+\sum_{i=2}^{n-1}\frac{\Delta_{i;\mu\nu}(p^{n-r})}{P_i}.
\end{equation}
For the first terms $\Gamma_{i;\mu\nu\rho}(p^{n-1-r})$, we require it to be gauge invariant in $n-r$ gluons, whereas the PMN of it is only $n-r-1$. \cite{Rodina:2016jyz} has shown that such terms should ruled out under this condition, then only the second terms survive in this mannar
\begin{equation}
    M_{n+1}^{(1)}(p^{n+1-r})=\epsilon_{n+1}^{[\mu}q^{\nu]}\sum_{i=2}^{n-1}\frac{\Delta_{i;\mu\nu}(p^{n-r})}{P_i}.
\end{equation}
It is obvious that this equation shares the same pattern with our discussion for the sub-leading terms, thus similarly, we obtain
\begin{equation}
    M_{n+1}^{(1)}(p^{n+1-r})=\left[\mathcal{A}_{n+1}^{\mathrm{YM}+\phi^3}(r)\right]^{(1)}.
\end{equation}

At the sub${}^{s\geqslant3}$-leading order, it is not straightforward to argue directly that $M_{n+1}^{(s-1)}(p^{n-r+1})=\left[\mathcal{A}_{n+1}^{\mathrm{YM}+\phi^3}(r)\right]^{(s-1)}$. Instead, we define
\begin{equation}
    B_{n+1}^{(s-1)}(p^{n-r+1}):=M_{n+1}^{(s-1)}(p^{n-r+1})-\left[\mathcal{A}_{n+1}^{\mathrm{YM}+\phi^3}(r)\right]^{(s-1)},
\end{equation}
and a alternative way is to show that $ B_{n+1}^{(s-1)}(p^{n-r+1})=0$ for $s\geqslant3$. Similarly, $ B_{n+1}^{(s-1)}(p^{n-r+1})$ has an expansion of 
\begin{equation}
    B_{n+1}^{(s-1)}(p^{n-r+1})=\epsilon_{n+1}^{[\mu}q^{\nu_1]}q^{\nu_2}\cdots q^{\nu_s}\left(\sum_{i=1,n}\frac{q^{\nu_{s+1}}\Gamma_{i;\mu\nu_1\dots\nu_{s+1}}(p^{n-r-s+1})}{q\cdot p_i}+\sum_{i=2}^{n-1}\frac{\Delta_{i;\mu\nu_1\dots\nu_s}(p^{n-r-s+2})}{P_i}\right).
\end{equation}
Since PMN of $\Delta_{i;\mu\nu_1\dots\nu_s}(p^{n-r-s+2})$ is less than $n-r$, there is no space for this factor to be gauge invariant in $n-r$ gluons, which leads to its nullification. Same goes with $\Gamma_{i;\mu\nu_1\dots\nu_{s+1}}(p^{n-r-s+1})$. Therefore we obtain that $B_{n+1}^{(s-1)}(p^{n-r+1})=0$, which tells us that
\begin{equation}
    M_{n+1}^{(s-1)}(p^{n-r+1})=\left[\mathcal{A}_{n+1}^{\mathrm{YM}+\phi^3}(r)\right]\qquad\text{for }s\geqslant3.
\end{equation}
Summarizing all these discussions, we show that for all order soft expansion, our anstaz can always be fixed by the $(n+1)$-point mixed amplitude: 
\begin{equation}
    M_{n+1}(p^{n-r+1})=\mathcal{A}_{n+1}^{\mathrm{YM}+\phi^3}(r).
\end{equation}
Such relation allows us to induct all the way down to the pure scalar amplitude $\mathcal{A}_r^{\phi^3}$. And for an $r$-scalar amplitude, it has $\mathcal{C}_{r-2}$ independent basis. This ends our proof.

\section{Details for 5-point planar universal expansion} \label{app: expansion}
The 5-point universal expansion reads:
\begin{equation}\label{eq: 5pt universal expansion}
     \begin{aligned}
        \mathcal{A} _{5}^{\mathrm{YM}}&=\epsilon _1\cdot \epsilon _5\mathcal{A} _{5}^{\mathrm{YM}+\phi ^3}\left( \left\{ 2,3,4 \right\} |1,5 \right)-\epsilon _1\cdot f_2\cdot \epsilon _5\mathcal{A} _{5}^{\mathrm{YM}+\phi ^3}\left( \left\{ 3,4 \right\} |1,2,5 \right)\\& -\epsilon _1\cdot f_3\cdot \epsilon _5\mathcal{A} _{5}^{\mathrm{YM}+\phi ^3}\left( \left\{ 2,4 \right\} |1,3,5 \right) -\epsilon _1\cdot f_4\cdot \epsilon _5\mathcal{A} _{5}^{\mathrm{YM}+\phi ^3}\left( \left\{ 2,3 \right\} |1,4,5 \right) 
\\ &
 +\epsilon _1\cdot f_2\cdot f_3\cdot \epsilon _5\mathcal{A} _{5}^{\mathrm{YM}+\phi ^3}\left( \left\{ 4 \right\} |1,2,3,5 \right) +\epsilon _1\cdot f_3\cdot f_2\cdot \epsilon _5\mathcal{A} _{5}^{\mathrm{YM}+\phi ^3}\left( \left\{ 4 \right\} |1,3,2,5 \right) 
\\
& +\epsilon _1\cdot f_2\cdot f_4\cdot \epsilon _5\mathcal{A} _{5}^{\mathrm{YM}+\phi ^3}\left( \left\{ 3 \right\} |1,2,4,5 \right) +\epsilon _1\cdot f_4\cdot f_2\cdot \epsilon _5\mathcal{A} _{5}^{\mathrm{YM}+\phi ^3}\left( \left\{ 3 \right\} |1,4,2,5 \right) 
\\
& +\epsilon _1\cdot f_3\cdot f_4\cdot \epsilon _5\mathcal{A} _{5}^{\mathrm{YM}+\phi ^3}\left( \left\{ 2 \right\} |1,3,4,5 \right) +\epsilon _1\cdot f_4\cdot f_3\cdot \epsilon _5\mathcal{A} _{5}^{\mathrm{YM}+\phi ^3}\left( \left\{ 2 \right\} |1,4,3,5 \right) 
\\
& -\epsilon _1\cdot f_2\cdot f_3\cdot f_4\cdot \epsilon _5\mathcal{A} _{5}^{\mathrm{YM}+\phi ^3}\left( \varnothing |1,2,3,4,5 \right) -\epsilon _1\cdot f_2\cdot f_4\cdot f_3\cdot \epsilon _5\mathcal{A} _{5}^{\mathrm{YM}+\phi ^3}\left( \varnothing |1,2,4,3,5 \right) 
\\
& -\epsilon _1\cdot f_3\cdot f_2\cdot f_4\cdot \epsilon _5\mathcal{A} _{5}^{\mathrm{YM}+\phi ^3}\left( \varnothing |1,3,2,4,5 \right) -\epsilon _1\cdot f_3\cdot f_4\cdot f_2\cdot \epsilon _5\mathcal{A} _{5}^{\mathrm{YM}+\phi ^3}\left( \varnothing |1,3,4,2,5 \right) 
\\
& -\epsilon _1\cdot f_4\cdot f_2\cdot f_3\cdot \epsilon _5\mathcal{A} _{5}^{\mathrm{YM}+\phi ^3}\left( \varnothing |1,4,2,3,5 \right) -\epsilon _1\cdot f_4\cdot f_3\cdot f_2\cdot \epsilon _5\mathcal{A} _{5}^{\mathrm{YM}+\phi ^3}\left( \varnothing |1,4,3,2,5 \right), 
    \end{aligned}
\end{equation}
According to~\eqref{eq: planar universal expansion} the original amplitudes are reorganized into:
\begin{align}
    \mathcal{A} _{5}^{\mathrm{YM}+\phi ^3}\left( \left\{ 2,3,4 \right\} |1,5 \right) &=\mathbf{A}_5^{\mathrm{YM}+\phi^3}\left( \left\{ 2,3,4 \right\} |15 \right) ,
\\
\mathcal{A} _{5}^{\mathrm{YM}+\phi ^3}\left( \left\{ 3,4 \right\} |1,2,5 \right)& =\mathbf{A}_5^{\mathrm{YM}+\phi^3}\left( \left\{ 3,4 \right\} |125 \right) ,  
\\
\mathcal{A} _{5}^{\mathrm{YM}+\phi ^3}\left( \left\{ 2,4 \right\} |1,3,5 \right) &=\mathbf{A}_5^{\mathrm{YM}+\phi^3}\left( \left\{ 2,4 \right\} |135 \right) ,  
\\
\mathcal{A} _{5}^{\mathrm{YM}+\phi ^3}\left( \left\{ 2,3 \right\} |1,4,5 \right) &=\mathbf{A}_5^{\mathrm{YM}+\phi^3}\left( \left\{ 2,3 \right\} |145 \right) ,
\\
\mathcal{A} _{5}^{\mathrm{YM}+\phi ^3}\left( \left\{ 4 \right\} |1,2,3,5 \right) &=\mathbf{A}_5^{\mathrm{YM}+\phi^3}\left( \left\{ 4 \right\} |1235 \right) +\mathbf{A}_5^{\mathrm{YM}+\phi^3}\left( \left\{ 4 \right\} |1[23] 5 \right) ,  
\\
\mathcal{A} _{5}^{\mathrm{YM}+\phi ^3}\left( \left\{ 4 \right\} |1,3,2,5 \right) &=-\mathbf{A}_5^{\mathrm{YM}+\phi^3}\left( \left\{ 4 \right\} |1[23] 5 \right) ,
\\
\mathcal{A} _{5}^{\mathrm{YM}+\phi ^3}\left( \left\{ 3 \right\} |1,2,4,5 \right) &=\mathbf{A}_5^{\mathrm{YM}+\phi^3}\left( \left\{ 3 \right\} |1245 \right) +\mathbf{A}_5^{\mathrm{YM}+\phi^3}\left( \left\{ 3 \right\} |1[24] 5 \right) ,  
\\
\mathcal{A} _{5}^{\mathrm{YM}+\phi ^3}\left( \left\{ 3 \right\} |1,4,2,5 \right) &=-\mathbf{A}_5^{\mathrm{YM}+\phi^3}\left( \left\{ 3 \right\} |1[24] 5 \right) ,
\\
\mathcal{A} _{5}^{\mathrm{YM}+\phi ^3}\left( \left\{ 2 \right\} |1,3,4,5 \right) &=\mathbf{A}_5^{\mathrm{YM}+\phi^3}\left( \left\{ 2 \right\} |1345 \right) +\mathbf{A}_5^{\mathrm{YM}+\phi^3}\left( \left\{ 2 \right\} |1[34] 5 \right) ,  
\\
\mathcal{A} _{5}^{\mathrm{YM}+\phi ^3}\left( \left\{ 2 \right\} |1,4,3,5 \right) &=-\mathbf{A}_5^{\mathrm{YM}+\phi^3}\left( \left\{ 2 \right\} |1[34] 5 \right) ,
\\
\mathcal{A} _{5}^{\mathrm{YM}+\phi ^3}\left( \varnothing |1,2,3,4,5 \right) &=\mathbf{A}_5^{\mathrm{YM}+\phi^3}\left( \varnothing |12345 \right) +\mathbf{A}_5^{\mathrm{YM}+\phi^3}\left( \varnothing |1[[23] 4] 5 \right) \\&+\mathbf{A}_5^{\mathrm{YM}+\phi^3}\left( \varnothing |12[34] 5 \right) +\mathbf{A}_5^{\mathrm{YM}+\phi^3}\left( \varnothing |1[23] 45 \right) \\
&+\mathbf{A}_5^{\mathrm{YM}+\phi^3}\left( \varnothing |1[2[34]] 5 \right) ,
\\
\mathcal{A} _{5}^{\mathrm{YM}+\phi ^3}\left( \varnothing |1,2,4,3,5 \right) &=-\mathbf{A}_5^{\mathrm{YM}+\phi^3}\left( \varnothing |12[34] 5 \right) -\mathbf{A}_5^{\mathrm{YM}+\phi^3}\left( \varnothing |1[2[34]] 5 \right) ,
\\
\mathcal{A} _{5}^{\mathrm{YM}+\phi ^3}\left( \varnothing |1,3,2,4,5 \right) &=-\mathbf{A}_5^{\mathrm{YM}+\phi^3}\left( \varnothing |1[23] 45 \right) -\mathbf{A}_5^{\mathrm{YM}+\phi^3}\left( \varnothing |1[[23] 4] 5 \right) ,
\\
\mathcal{A} _{5}^{\mathrm{YM}+\phi ^3}\left( \varnothing |1,3,4,2,5 \right) &=\mathbf{A}_5^{\mathrm{YM}+\phi^3}\left( \varnothing |1[2[34]] 5 \right) ,
\\
\mathcal{A} _{5}^{\mathrm{YM}+\phi ^3}\left( \varnothing |1,4,2,3,5 \right) &=\mathbf{A}_5^{\mathrm{YM}+\phi^3}\left( \varnothing |1[[23] 4] 5 \right) ,
\\
\mathcal{A} _{5}^{\mathrm{YM}+\phi ^3}\left( \varnothing |1,4,3,2,5 \right) &=-\mathbf{A}_5^{\mathrm{YM}+\phi^3}\left( \varnothing |1[[23] 4] 5 \right) -\mathbf{A}_5^{\mathrm{YM}+\phi^3}\left( \varnothing |1[2[34]] 5 \right) ,
\end{align}
which leads to~\eqref{eq: 5pt planar universal expansion}.

\newpage
  \bibliographystyle{JHEP}
  \bibliography{Refs}
\end{document}